\documentclass[12pt, a4]{article}

\usepackage[mathscr]{euscript}
\usepackage[makeroom]{cancel}
\usepackage[font={small,it,singlespacing}]{caption}
\usepackage[colorlinks = true, linkcolor = red, urlcolor = blue, citecolor = blue, anchorcolor = green]{hyperref}
\usepackage{amsmath, amssymb, amsthm, amsbsy, amsfonts, bigints}
\usepackage{enumerate, verbatim, sectsty, colortbl, tabularx,enumitem, algorithmic}
\usepackage[linesnumbered,ruled,vlined]{algorithm2e}
\usepackage{natbib}
\usepackage{graphicx, float, pdflscape,rotating, pspicture, epsfig, subcaption, threeparttable, multirow, multicol, afterpage,lscape}
\usepackage{setspace}
\usepackage[compact]{titlesec}
\usepackage[margin=1in]{geometry}
\usepackage[belowskip=2pt,aboveskip=2pt]{caption}

\allowdisplaybreaks

\newcommand{\bs}{\boldsymbol}
\newcommand{\R}{\mathcal{R}}

\newcommand{\x}{\mathbf{x}}
\newcommand{\X}{\mathbf{X}}
\newcommand{\y}{\mathbf{y}}
\newcommand{\s}{\mathbf{s}}
\newcommand{\e}{\mathbf{e}}

\newcommand{\E}{\mbox{E}}
\newcommand{\V}{\mbox{V}}

\theoremstyle{definition}
\newtheorem{defn}{Definition}
\newtheorem{thm}{Theorem}

\newtheorem{pro}[thm]{Proposition}
\newtheorem{rem}[thm]{Remark}

\makeatletter
\g@addto@macro\normalsize{%
  \setlength{\abovedisplayskip}{3pt plus 2pt}%
  \setlength{\belowdisplayskip}{3pt plus 2pt}%
  \setlength{\abovedisplayshortskip}{3pt plus 2pt}%
  \setlength{\belowdisplayshortskip}{3pt plus 2pt}%
  \setlength{\belowcaptionskip}{-3pt}
}
\makeatother

\begin{document}
\title{\large{\textbf{Model-based Differentially Private Data Synthesis and Statistical Inference in Multiply Synthetic Differentially Private Data \vspace{-9pt}}}}
\author{\small{\textbf{Fang Liu\footnote{Fang Liu is Professor in the Department of Applied and Computational Mathematics and Statistics, University of Notre Dame, Notre Dame, IN 46556 ($^{\ddag}$E-mail: fang.liu.131@nd.edu). The work was supported by the NSF Grant  1546373 and the University of Notre Dame Faculty Research Initiation Grant.}}}\\
\small{Department of Applied and Computational Mathematics and Statistics} \\
\small{University of Notre Dame, Notre Dame, IN 46556}\\
\small{fang.liu.131@nd.edu}
}
\date{} %\small{\today}}

\maketitle
\vspace{-21pt}
\begin{abstract}
\noindent We propose the approach of model-based differentially private synthesis (modips) in the Bayesian framework for releasing individual-level surrogate/synthetic datasets with privacy guarantees given the original data.  The modips technique integrates the concept of differential privacy into model-based data synthesis. We introduce several variants for the general modips approach and different procedures to obtaining privacy-preserving posterior samples, a key step in modips. The uncertainty from the sanitization and synthetic process in modips can be accounted for by releasing multiple synthetic datasets and quantified via an inferential combination rule that is proposed in this paper.  We run empirical studies to examine the impacts of the number of synthetic sets and the privacy budget allocation schemes on the inference  based on synthetic data.

\vspace{3pt }\noindent  \textit{\textbf{keywords}}:  (Bayesian) sufficient statistics,  budget allocation, differentially private posterior distribution, inference,  multiple release, sanitization,  surrogate data
\end{abstract}

\setstretch{1.03}
\vspace{-9pt}
\section{Introduction}\label{sec:introduction}\vspace{-9pt}
\subsection{Background and Motivation}\vspace{-6pt}
Data synthesis (DS) is a statistical disclosure limitation  technique that releases pseudo  individual-level data for research and public use.  Both parametric and nonparametric  Bayesian and frequentist approaches have been proposed for DS \citep{rubin1993statistical, Liu2002, Liu2004, ReiterCART2005, an2007, ReiterRandomForest2010, rdrechsler2011paper, Burgette2013}. To  reflect the uncertainty introduced during the synthesis process, multiple sets of synthetic data are often released. Inferential methods  that  combine information from multiple synthetic datasets to yield valid inferences  are available \citep{raghunathan2003multiple,reiter2003}.  A  long-standing research problem in statistical disclosure limitation is the lack of a universally applicable and robust measure of disclosure risk in released data. Many existing disclosure risk assessment approaches rely on strong and ad-hoc assumptions on the  background knowledge and behaviors of data intruders \citep{fienberg1997bayesian, Domingo2001, Domingo2004, reiter2005estimating, Domingobook2008, manrique2012estimating,  reiter2014bayesian}.

Differential privacy (DP) has gained enormous popularity since its debut in 2006 \citep{dwork2006calibrating}. DP formalizes privacy in mathematical terms without making  assumptions about data intruders and has nice properties such as privacy cost composition and immunity to post processing for the information sanitized through a DP randomization algorithm. DP has spurred a great amount of work in developing differentially private randomization mechanisms to release statistics in general settings as well as for specific types of queries or statistical analyses. The Laplace mechanism \citep{dwork2006calibrating}, the Exponential mechanism \citep{mcsherry2007mechanism}, and  the Gaussian mechanism \citep{privacybook, ggm} are common differentially private sanitizers for general purposes.  Differentially private versions of various statistical analyses are also available, such as point estimators \citep{estimation, m-estimator}, principle components analysis \citep{Chaudhuri2012PCA},    linear and penalized regression \citep{Chaudhuri2011, Kifer2012}, model selection \citep{lasso}, release of functions \citep{function}, the $\chi^2$ test in genome-wide association studies \citep{Yu2014}, and deep learning \citep{deep1, deep2}, among others.

In addition to the sanitization of aggregates statistics, there is also \underline{di}fferentially  \underline{p}rivate  \underline{s}ynthesis of individual-level data (dips). An obvious advantage of dips over query-based sanitization is that it releases surrogate datasets of the same structure as the original dataset that allows data users to run analyses of their own as if they had the original data.  Dips also eliminates the need to continuously monitor submitted queries and design differentially private algorithms to sanitize query results in interactive settings, offering a practical solution considering that it is unlikely for data curators to anticipate  the types of queries submitted to a database  beforehand. In addition,  the total privacy budget is often fixed, allowing only a limited number of queries to be answered before the budget is exhausted per the sequential composability  of DP \citep{mcsherry2007mechanism}.

\vspace{-3pt}\subsection{Related work}\vspace{-9pt}\label{sec:work}
Early dips approaches are model-free in nature or focus on categorical data synthesis and requires some discretization for numerical attributes. In the framework of discrete data synthesis via model-free approaches, \citet{barak2007privacy} generated synthetic data via the Fourier transformation and linear programming  in low-order contingency tables. \citet{blum2008learning} discussed the possibility of dips from the perspective of the learning theory in a discretized domain. \citet{multiplicative} developed the iterative MWEM (multiplicative weights exponential mechanism) algorithm to synthesize discrete data via ``matching'' on linear queries.  \citet{bowen2021differentially} propose the STEPS procedure that partitions data by attributes  according to  a practical or statistical importance measure and synthesize the data from the constructed hierarchical attribute tree.  \citet{eugenio2021construction} propose the CIPHER procedure to construct differentially private empirical Distributions from a set of low-order marginals through solving linear equations with $l_2$ regularization.

For model-based categorical data synthesis, the multinomial-Dirichlet synthesizer  \citep{abowd2008protective} designs a prior for cell proportions to achieves DP for tabular data in the Bayesian framework. The approach was applied to synthesize the US commuting data (\url{https://onthemap.ces.census.gov/}) in \citet{onthemap} and its inferential properties were studied in \citet{Charest2010}. \citet{mcclure2012differential} implemented a similar concept but with a different prior for the binomial-beta model to synthesize univariate binary data.   \citet{privbayes} proposed PrivBayes to release high-dimensional categorical data from Bayesian networks. Bayesian network is a probabilistic graphical model and does not involve Bayesian modeling or inference; thus PrivBayes conceptually differs from the Bayesian dips framework that we focus on.  %\footnote{Bayesian network is a probabilistic graphical model but does not involve Bayesian modelling or inference; thus PrivBayes conceptually differs from the Bayesian dips framework that we focus on.}  

For numerical data synthesis, \citet{wasserman2010statistical} proposed several paradigms to sample from differentially private perturbed histograms, smoothed histograms, and cumulative distribution functions; and examined the rate at which the differentially private distributions converge to the true distribution. \citet{function} proposed a differentially private kernel density estimator. In both works,  synthetic data can be simulated and released from the privacy-preserving density functions.  \citet{copula} proposed DPCopula to sample synthetic data from differentially private copula functions for multi-dimensional data.  \citet{quick2019generating} proposed an approach to generate private synthetic data via the Poisson-gamma model and applied the approach to disease mapping.  %DualQuery by \citet{dualquery} also employed the MW technique to handle a large number of linear queries  in the discrete domain.  
There are also dips approaches for specific types of data such as graphs \citep{graph,wang2013preserving,xiao2014differentially,li2016differentially}; mobility data from GPS trajectories \citep{andres2013geo, he2015}, and edge data based on exponential random-graph models in social networks \citep{jrssc,liu2020differentially}.

Some recent work employs neural networks (NN), including Generative Adversarial Networks (GAN) \citep{goodfellow2014generative}, to release synthetic data from differentially private generative models \citep{acs2018differentially,kang2020study,abay2018privacy,jordon2018pate}. One of the advantages of the NN-based approach is that it relies on machine learning methods to develop a robust generative model and does not make distributional assumptions on the training data. On the other hand, NNs are often subject to overfitting and the robustness of a trained NN often replies on large training data sizes. \cite{bindschaedler2017plausible} design a framework to achieve plausible deniability instead of achieving PD directly and use a privacy test to reject ``bad'' samples that do not satisfy plausible deniability. Though this method does not directly inject noises into the generative model, it is very likely this rejection-of-bad-sample step can systematically bias the synthetic data, leading to invalid inference based on this intentionally selected ``safe'' data.  Dips is also a topic for  doctoral dissertation research in recent years \citep{bindschadler2018privacy, melis2018building,bowen,eugenio}.

\subsection{Our Contributions}\vspace{-9pt}
We aim to develop a general framework that integrates the notion of DP into data synthesis to achieve formal privacy guarantees. Toward that end, we develop the   \underline{mo}del-based \underline{di}fferential \underline{p}rivate \underline{s}ynthesis (modips) framework in the Bayesian framework. Different from the existing Bayesian dips approaches such as the multinomial-Dirichlet and the beta-binomial synthesizers, the modips does not achieve DP through prior specification, but rather through sanitizing the posterior distribution. modips is a general approach that can handle all data types (numerical, categorical, discrete, relational) where an appropriate Bayesian model can be constructed. Our main contributions are summarized as follows. \vspace{-9pt}
\begin{itemize}[leftmargin=12pt,itemsep=-2pt]
\item We achieve DP in the modips procedure in the step of sanitizing the posterior distribution of the Bayesian model parameters. We propose several specific procedures to obtaining privacy-preserving posterior samples in this step. The achieved privacy guarantees are preserved in the subsequent synthesis steps and in released surrogate datasets. 
\item We recommend releasing multiple synthetic datasets so that the  uncertainty of the sanitization and synthesis processes can be conveniently accounted for in inference from the sanitized data.  We examine the asymptotic properties of the inference and provide an inferential combination rule across multiple synthetic datasets. We run empirical studies to examine the impact of the number of multiple synthetic sets on inference.
\item We propose the concepts of communal sanitization and individualized sanitization and study the effect of privacy budget allocation on inference in the individualized sanitization via an empirical study.  
\end{itemize} \vspace{-9pt}
The rest of the paper is organized as follows. Sec \ref{sec:prelim} overviews the basic concepts of DP, some commonly used differentially private mechanisms for releasing information, and the traditional non-DP-based multiple synthesis procedure.  Sec \ref{sec:dips} introduces the modips approach and several specific procedures to obtaining privacy-preserving posterior samples, a key step in the modips approach. Sec \ref{sec:inference} proposes an approach to combine information from multiply differentially private synthetic datasets for valid inference. Sec \ref{sec:examples} runs several simulation studies to illustrate the application of  modips, validate the inferential combination rule and examine the effects of the number of released datasets and the impact of privacy budget allocation  on inferences based on the sanitized data; it also summarizes the results from published work that implements the modips approach or uses our inferential combination rule.  The paper concludes in Sec \ref{sec:discussion} with final remarks and some topics for future work.

%%%%%%%%%%%%%%%%%%%%%%%%%%%%%%%%%%%%%%%%%%%%%%%%%%%%%%%%%%%%%%%%%%%%%%%%%%%%
\section{Preliminaries}\label{sec:prelim}
%%%%%%%%%%%%%%%%%%%%%%%%%%%%%%%%%%%%%%%%%%%%%%%%%%%%%%%%%%%%%%%%%%%%%%%%%%%%
\vspace{-9pt}\subsection{Differential privacy}\vspace{-6pt}
\begin{defn} \citep{dwork2006, dwork2006calibrating} A sanitization algorithm $\R$ is $\epsilon$-differentially private if,  for all datasets $(\x,\x')$ that differ in one individual and all possible subset $Q$ to the output range of statistics $\s$ from $\R$,
%\begin{equation}\label{eqn:dp}
$\left|\log\left(\frac{\Pr(\R(\s,\x)) \in Q)}{\Pr(\R(\s,\x'))\in Q)} \right)\right|\le\epsilon$.
%\end{equation}
\vspace{-9pt}\end{defn}\label{def:dp}
\noindent  $\epsilon>0$ is the privacy loss or budget parameter.  The DP definition implies that the probabilities of  obtaining the same statistic from $\x$ and $\x'$ after the sanitization  are similar -- the ratio between the two probabilities falls within $\in[e^{-\epsilon}, e^{\epsilon}]$. In layman's terms, DP implies the chance that an individual in a dataset will be identified based on the released sanitized  results is very low since the results are about the same with or without that individual. The smaller $\epsilon$ is, the more protection will be executed on the individuals in the dataset. 

In what follows, we use $d(\x,\x')=1$ to denote two datasets $\x,\x'$ differing by  one individual, which is defined in two ways. First, $\x,\x'$  have the same sample size, but one and only one record  differs in at least one attribute; a substitution would make $\x,\x'$  identical(aka \emph{bounded DP} in \citet{kifer2011no}). In the second definition, one dataset has one more record than the other, so the sample sizes differ by 1, and deletion (or insertion) of one record would make $\x,\x'$ identical (aka ``unbounded  DP'').

DP provides strong and robust privacy guarantees in the sense that it does not make  assumptions regarding the background knowledge or behaviors of data intruders. In some practical applications,  satisfying the pure DP in Def \ref{def:dp} might lead to much perturbation/sanitization in released information. To lessen the degree of perturbation, various softer versions of DP have been developed, such as the  $(\epsilon,\delta)$-approximate DP (aDP) \citep{dwork2006delta}, the $(\epsilon,\delta)$-probabilistic DP (pDP) \citep{onthemap}, the $(\epsilon,\delta)$-random DP (rDP) \citep{randomDP}, the $(\epsilon,\tau)$-concentrated DP (cDP) \citep{cPD} and truncated cDP \citep{bun2018composable}, R{\'e}nyi DP \citep{mironov2017renyi}, and the most recent  Gaussian DP \citep{dong2019gaussian}.  In many relaxed versions of DP,  extra parameters are employed  to characterize the amount of relaxation on top of the privacy budget $\epsilon$ and include the  pure DP as a special case. For example, $(\epsilon,\delta)$-aDP and  $(\epsilon,\delta)$-pDP reduce to $\epsilon$-DP when $\delta=0$, and $(\alpha, \epsilon$-R{\'e}nyi DP reduces to $\epsilon$-DP when $\alpha=\infty$. The Gaussian DP uses a functional relaxation to replace explicit privacy loss parameters.

Many differentially private mechanisms have been proposed to sanitize statistics, among which the Laplace mechanism, the Gaussian mechanism,  and the exponential mechanism are three popular sanitizers of $\epsilon$-DP for general purposes. 
\vspace{-3pt}\begin{defn}\citep{dwork2006calibrating} Let $\s=(s_1,\ldots,s_r)$ be a $r$-dimensional statistic and $\mathbf{e}$ comprise $r$ independent random samples from Laplace$\left(0,\Delta_1\epsilon^{-1}\right)$ and $\Delta_1$ is the $l_1$ global sensitivity of $\s$. The sanitized $\s^*$ via the \emph{Laplace mechanism} of  $\epsilon$-DP is $\s^*=\s+\mathbf{e}$.\vspace{-12pt}
\end{defn}
%=\!\!\!\underset{\scriptstyle{\x,\x', d(\x,\x')=1}}{\mbox{max}} (\;\textstyle\sum_{i=1}^r|\s(\x)-\s(\x')\|^p)^{1/p}$ 
$\Delta_1$ is a special case of the $l_p$ \emph{global sensitivity} (GS) $\Delta_p=\mbox{max}_{\scriptstyle{\x,\x', d(\x,\x')=1}} \|\s(\x)-\s(\x')\|_p$ when $p=1$ \citep{ggm}. The sensitivity is ``global" since it is defined for all possible datasets and all possible ways of two neighboring datasets differing by one record. The larger the GS for $\s$ is, the larger the privacy risk is from releasing the original $\s$, and the more perturbation is needed for $\s$ to offset the large sensitivity. This is also reflected in the variance of the Laplace distribution $2\left(\delta_1\epsilon^{-1}\right)^2$: the larger $\delta_1$ or the smaller $\epsilon$ is, the more spread the distribution of $\s^*$ is, and the more likely that extreme $\s^*$ values that are far away from $\s$ will be released. 
\vspace{-3pt}\begin{defn}\citep{dwork2014algorithmic,ggm} \vspace{-3pt}
Let $\s\!=\!(s_1,\ldots,s_r)$ be a $r$-dimensional statistic. The \emph{Gaussian mechanism} sanitizes $\s$ as in $\s^*=\s+\e$, where $e_j\sim \mathcal{N}(0,\sigma^2)$ independently for $j=1,\ldots,r$ with $\sigma \geq c\Delta_2/ \epsilon$ ($\Delta_2$ is the $l_2$ GS of $\s$) for $\epsilon<1$ and  $c^2>2\log(1.25/\delta)$ in the case of $(\epsilon,\delta)$-aDP and
$\sigma\!\geq\!(2\epsilon)^{\!-1} \Delta_2(\!\sqrt{(\Phi^{-1}(\delta/2))^2+2\epsilon }\!-\!\Phi^{-1}(\delta/2))$
in the case of $(\epsilon,\delta)$-pDP, where $\Phi^{-1}$ is the inverse CDF of the standard normal distribution.
\end{defn}\vspace{-6pt}
\begin{defn}\citep{mcsherry2007mechanism} Let $\s=(s_1,\ldots,s_r)$ be a $r$-dimensional statistic and $\mathcal{S}$ be the set containing all possible sanitized outputs. The \emph{exponential mechanism} of $\epsilon$-DP releases $\s^{\ast}$  from
$p(\s^{\ast}|\x)\!=\!\frac{\exp\left(u(\s^{\ast}|\x)\epsilon/(2\Delta_u)\right)} {\sum_{\s^{\ast}\in\mathcal{S}} \exp\left(\!u(\s^{\ast}|\x)\epsilon/(2\Delta_u)\!\right)}$ for discrete outputs and from
$p(\s^{\ast}|\x)\!=\!\frac{\exp\left(u(\s^{\ast}|\x)\epsilon/(2\Delta_u)\right)}{\int_{\s^{\ast}\in\mathcal{S}} \exp\left(u(\s^{\ast}|\x)\epsilon/(2\Delta_u)\!\right)}$ for continuous outputs, where $u(\s^*|\x)$ is the utility score of $\s^*$ given data $\x$ and  $\delta_u=\max_{\scriptstyle{\x,\x', \delta(\x,\x')=1}} |u(\s^{\ast}|\x)-u(\s^{\ast}|\x')|$ is the maximum change (sensitivity) in score $u$ for all pairs of neighboring datasets $\x$ and $\x'$. \vspace{-9pt}
\end{defn}

DP has some nice properties that other privacy notions do not possess. One of such properties is that sanitized results through DP mechanisms are immune to post-processing in that the results do not leak more private information about the individuals if they are further processed after release (as long as there is no access to the original data from which the results are calculated). Another nice property of the pure DP and most of its relaxed forms is that the privacy loss from applying multiple differentially private mechanisms to the same dataset is closed under composition (note this does not apply to some of the relaxed DP definitions; see \citet{kifer2012axiomatic}). There are two  basis composition principles in DP: \emph{parallel composition} and \emph{sequential composition} \citep{mcsherry2007mechanism}. If mechanism $\R_j$ is $\epsilon_j$-DP for $j=1,\dots,r$ and each is applied on disjoint subsets $D_j$ of a dataset $D$, then $\prod_j\R_j(\y\cap D_j)$ is $\max{(\epsilon_j)}$-DP per the parallel composition; if $\R_j$ is applied to the same dataset $D$, then $\prod_j\R_j(\y)$ is $(\sum_j\epsilon_j)$-DP per the sequential composition. Besides the basic composition, there is also advanced composition \citep{dwork2010boosting} that provides tighter privacy loss bounds. Some relaxed DP definitions allow the privacy loss composition to be handled exactly and easily tracked (e.g., cDP, truncated cDP, R{\'e}nyi DP).

\vspace{-3pt}\subsection{Multiple Synthesis}\vspace{-9pt}
The surrogate datasets generated through multiple synthesis (MS) have the same structure as the original but contain pseudo-individuals synthesized in a model-free or model-based framework given the original data.  Depending on the data source that the  synthesis is based on, the traditional MS approaches can be roughly grouped into population synthesis and sampling \citep{rubin1993statistical,raghunathan2003multiple} and sample synthesis \citep{little1993}. By the percentage of the synthetic component in a released dataset, DS can be grouped into partial synthesis and full synthesis \citep{Liu2002, Liu2004}.  

\begin{figure}[!htb]\begin{center}\vspace{-6pt}
\includegraphics[scale=0.75]{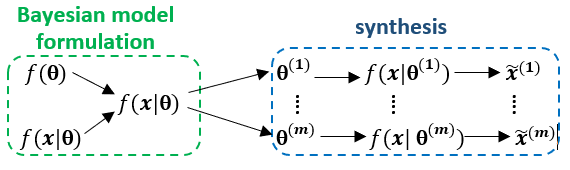}
\caption{The traditional MS procedure}\label{fig:MS}\vspace{-24pt}
\end{center}\end{figure}

Fig \ref{fig:MS} depicts a Bayesian MS procedure sample full synthesis, the framework that we focus on in this paper. In brief, a Bayesian model is first formulated given data $\x$; then multiple  ($m$) sets of model parameters $\bs\theta$ are drawn from the posterior distributions; finally, $m$ sets of surrogate datasets ($\x^{(1)},\ldots,\x^{(m)}$) are generated, one for each $\bs\theta$.

The MS procedure in Fig \ref{fig:MS} does not use external randomization algorithms to sanitize the information in the original $\x$. The arguments for privacy guarantees are often heuristic as no individuals in the surrogate data correspond to any real persons. In recent years, there is work that connects posterior sampling, which has inherent randomness, with the DP concept. \cite{wang2015privacy} proved that the privacy loss for releasing one posterior sample of  $\bs{\theta}$ given any prior is $4B$, where $B$ is the upper bound of the log-likelihood $\log(l(\bs{\theta}|\x)$. \cite{dimitrakakis2014robust} show that if the change in the log-likelihood between two neighboring datasets ($\x,\x'$) is bounded by a constant $C$, releasing one  posterior sample of $\bs{\theta}$ is $2C$-differentially private. 
For MS, since $m>1$ posterior samples are released, the overall privacy loss increases $m$ folds and becomes $4mB$ and $2mC$, respectively. In brief, the privacy loss for releasing one posterior sample depends on the inherent properties of the likelihood function rather than a parameter can be specified. If the bounds ($B$ and $C$) are large, the privacy loss from releasing a posterior sample can be too large to provide sufficient privacy guarantees. 

%%%%%%%%%%%%%%%%%%%%%%%%%%%%%%%%%%%%%%%%%%%%%%%%%%%%%%%%%%%%%%%%%%%%%%%%%%%

\section{\large{Model-based Differentially Private Data Synthesis (modips)}}\label{sec:dips}\vspace{-9pt}
\subsection{The modips Procedure}
\begin{figure}[!htb]\vspace{-12pt}\begin{center}
\includegraphics[scale=0.75]{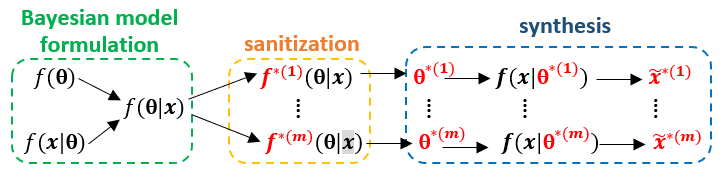}\vspace{-3pt}
\caption{The modips Procedure. $f(\x|\bs{\theta})$ is the model assumed on the original data $\x$, $f(\bs{\theta})$ is the prior, and $f(\bs{\theta}|\x)$ is the posterior distribution of $\bs{\theta}$; the superscript * represents ``sanitized''}\label{fig:modips1}
\end{center}\vspace{-24pt}\end{figure}

\vspace{-9pt}
Fig \ref{fig:modips1} presents an illustration  diagram of the modips procedure. The procedure comprises 3 sequential steps: model formulation, sanitization, and data synthesis. Its output is $m\ge1$ sets of synthetic data, each of the same data structure as the original dataset $\x$. 

The algorithmic steps of  modips are presented in Algorithm \ref{alg:modips1}. With $m$ released datasets, each set is sanitized with $1/m$ of the overall privacy budget per the sequential composition. Since the amount of noise increases with decreasing privacy budget, this implies a synthetic set for $m\!>\!1$ is noisier than that for $m\!=\!1$. However, the totality of released original information across the $m$ released sets for $m\!>\!1$  is not necessarily less than that at $m\!=\!1$. More importantly, releasing multiple sets provides an effective and convenient way to quantify

\begin{algorithm}[H]
\caption{The modips Procedure}\label{alg:modips1}
\SetAlgoLined
\SetKwInOut{Input}{input}
\SetKwInOut{Output}{output}
\Input{number of synthetic datasets $m$, privacy budget $\epsilon$, original data $\x$, Bayesian model set $\mathcal{M}$ } 
\Output{surrogate datasets:  $\tilde{\x}^{*(1)},\ldots, \tilde{\x}^{*(m)}$} 
If $|\mathcal{M}|>1$, select a Bayesian model from set $\mathcal{M}$ via the exponential mechanism with budget $\epsilon_0<\epsilon$; else, set $\epsilon_0=0$ and got to line 2\;
\For{$i=1,\ldots,m$}{
Obtain a posterior sample $\bs\theta^{*(i)}$ from  $f(\bs{\theta}|\x)$ corresponding to the selected Bayesian model via  a  differentially private mechanism with privacy budget $(\epsilon-\epsilon_0)/m$\;
Draw $\tilde{\x}^{*(i)}$ from $f\big(\x|\bs{\theta} ^{*(i)}\big)$.
}
\end{algorithm}

the uncertainty and randomness introduced during the sanitization and synthesis that is necessary for valid inferences given the released data if no other sources or approaches are available to data users to quantify the uncertainty (see Sec \ref{sec:inference} for details). 

Algorithm \ref{alg:modips1} starts with a model selection step  if the user does not have a pre-specified model for the data $\x$. Since the model selection uses the information in $\x$, it costs privacy.  The utility function $u$ in the exponential mechanism can use metrics measuring model fitting on $\x$, such as the negative deviance information criterion (DIC). The probability that a model is selected from the candidate set $\mathcal{M}$ is proportional to $u$, calibrated simultaneously to the privacy budget $\epsilon_0$ assigned to the model selection step and the sensitivity of $u$. One can also incorporate the model selection step as part of the ``for'' loop; that is, each synthetic dataset is based on a different model selected via the exponential mechanism from $\mathcal{M}$. If the option is adopted, then the privacy budget for each model selection is $\epsilon_0/m$. There are pros or cons to this approach. On one hand, statistical inference based on the synthetic data $\tilde{\x}^{*(1)},\ldots, \tilde{\x}^{*(m)}$ ought to be more robust as it is implicitly averaged over multiple synthesis models. On the other hand, the inference is subject to more variability with the employment of multiple synthesis models, especially considering that the model selection privacy budget $\epsilon_0$ is further split into $m$ portions, making the model selection less meaningful from a utility preservation perspective. In practice, it often exists prior knowledge to help determine a model   without using the information in data $\x$. Therefore,  $|\mathcal{M}|=1$ and the model selection step can be skipped and all the privacy budget can be used toward the synthesis step. 

We provide several approaches to obtaining  posterior samples for $\bs\theta$ in a differentially private manner (line 3 in  Algorithm \ref{alg:modips1}) in Sec \ref{sec:sample}. Algorithm \ref{alg:modips1} is also applicable to relaxed versions of DP. For example, if $(\epsilon,\delta)$-aDP is employed, the data curator will specify values for both $\epsilon$ and $\delta$  and  split both between the model selection and synthesis steps and across the $m$ syntheses.
 
\begin{figure}[!htb]\vspace{-3pt}\begin{center}
\includegraphics[scale=0.75]{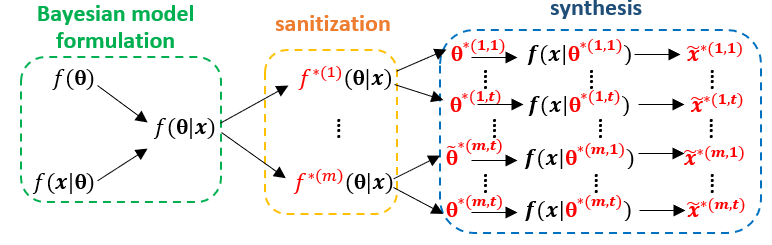}\vspace{-3pt}
\caption{The nested modips procedure}\label{fig:modips3}\vspace{-24pt}
\end{center}\end{figure}

A variant to the ``standard'' modips procedure (Fig \ref{fig:modips1}) is the  \emph{nested modips} procedure, as illustrated in Fig \ref{fig:modips3}. In brief, for a given $i=1,\ldots,m$, $t>1$ sets of $\bs{\theta}^{*(i,1)},\ldots, \bs{\theta}^{*(i,t)}$ are sampled, each of which leads to a synthetic dataset. The released $m\times t$ sets of surrogate data $\tilde{\x}^{*(1,1)},\ldots,\tilde{\x}^{*(1,t)},\ldots,\tilde{\x}^{*(m,1)},\ldots,\tilde{\x}^{*(m,t)}$ takes a 2-layer hierarchical structure. Since the output volume from the nested modips is $t$ folds of that of the standard modips procedure and the analysis of the synthetic data is also more complex with the hierarchical data structure, we suggest not employing the nested modips unless there is an absolute need or interest to separately quantify due to sanitization vs. synthesis (see Sec \ref{sec:inference} for detail).

\subsection{Privacy Guarantees of modips}\label{sec:dp}\vspace{-3pt}
\begin{pro}\label{pro:dp}\vspace{-3pt}
The modips procedure in Algorithm \ref{alg:modips1} is $\epsilon$-differentially private.
\end{pro}\vspace{-9pt}
The proof of Proposition \ref{pro:dp} is provided in the Appendix. The proof suggests that the step of drawing $\tilde{\x}^*$ is does not  incur any additional privacy cost as it can be regarded as post-processing of the already-sanitized $\bs\theta^*$ information.  The conclusion in  Proposition \ref{pro:dp} can be easily extended to a relaxed version of DP that is immune to post-processing and closed under composition, such as $(\epsilon,\delta)$-aDP. 

\vspace{-3pt}\subsection{Differentially Private Posterior Sampling}\label{sec:sample} \vspace{-6pt}
We present a few approaches to obtaining sanitized samples from $\!f(\bs{\theta}|\x)$ (line 3 of Algorithm \ref{alg:modips1}), including direct sanitization, sanitization through Bayesian sufficiency, and sanitization of approximate distribution. We introduce each approach in detail below.

\emph{Direct Sanitization}. One can directly sanitize the posterior distribution function $f(\bs{\theta}|\x)$ via a DP mechanism.  Though this sounds straightforward conceptually,  it can be difficult to implement practically. One of the reasons for the difficulty is that $f(\bs{\theta}|\x)$ is often only known up to a constant in many practical problems; that is, $f(\bs{\theta}|\x)\propto f(\x|\bs{\theta})f(\bs{\theta})$ and the normalizing constant $f(\x)$ might not have a close-form expression. This matters in the framework of DP as $f(\x)$ is a function of data  $\x$ -- the target for protection. Even if $f(\bs{\theta}|\x)$ has a closed form, sanitizing  $f(\bs{\theta}|\x)$ is not as simple as  $f^*(\bs{\theta}|\x)=f(\bs{\theta}|\x)+e$, say  $e\sim$ Lap$(0,\Delta_f\epsilon^{-1})$,  as $f^*(\bs{\theta}|\x)$ might no longer integrate into 1 or be a proper density function.  In fact, there are not many approaches in the literature to sanitizing density functions or CDFs. \citet{wasserman2010statistical} suggested employing the exponential mechanism to release CDFs. This approach requires specification of a  set of candidate CDFs and a scoring function that measures the utility of each candidate CDF. In order to maintain a certain level of utility in the sanitized CDF, the set  may need to be very large, implying potentially high computational costs and even practically infeasibility especially when $\bs\theta$ is high-dimensional. 

\begin{figure}[!htb]\vspace{-9pt}\begin{center}
\includegraphics[scale=0.725]{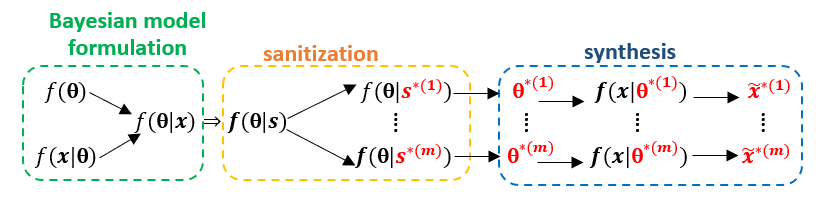}
\caption{The modips.SBS Procedure}\label{fig:modips2}
\end{center}\vspace{-18pt}\end{figure}

\emph{Sanitization through Bayesian Sufficiency (SBS)}. When $f(\bs{\theta}|\x)$ can be reformulated as  $f(\bs{\theta}|\s)$ where  $\s$ is a Bayesian sufficient statistic (scalar or multi-dimensional), one can sanitize $\s$ to achieve DP guarantees for $f(\bs{\theta}|\s)$ and thus for $f(\bs{\theta}|\x)$ and $f(\tilde{\x}|\bs{\theta})$. It is expected that the sanitization for modips.SBS is easier compared to the direct santization as $\s$ is of finite dimension and there are many existing mechanisms that can be used to sanitize statistics. We refer to this variant of the  modips procedure as modips.SBS (Fig \ref{fig:modips2}).  The formal privacy guarantees of modips.SBS is given in Proposition \ref{pro.SBS}; the proof can be found in the Appendix.  
\begin{pro}\label{pro.SBS}\vspace{-6pt}
The modips.SBS procedure satisfies $\epsilon$-DP.\vspace{-12pt}
\end{pro} 

In term of the actual implementation of modips.SBS, the steps are similar to Algorithm \ref{alg:modips1}, except for line 3; one would sanitize $\s$ with privacy budget $(\epsilon-\epsilon_0)/m$, say via the Laplace mechanism, to obtain $\s^{*(i)}$, and then  draw a posterior sample $\bs\theta^{*(i)}$ from  $f(\bs{\theta}|\s^{*(i)})$.  Identification of the sufficient statistic $\s$ in a Bayesian model is critical for the implementation of modips.SBS. Generally speaking, classical sufficiency implies Bayesian sufficiency.  There are examples of Bayesian sufficient statistics which are not classically sufficient but those are unusual situations \citep{Blackwell1982, Bernardo1994, Nogales2000}. More generally, even without assuming a parametric model, if $\s$ is predictive sufficient, that is, $\Pr(\tilde{\mathbf{X}}=\tilde{\x}\mid \mathbf{X}=\x)=\Pr(\tilde{\mathbf{X}}=\tilde{\x}\mid s(\mathbf{X})=s(\x))$, we can extend modips to a model-free framework by drawing $\tilde{\x}^{(i)*}$ directly from $f(\tilde{\x}^{(i)*}|\s^{(i)*})$  if this distribution is easy to compute and sample from. 

\emph{Sanitization of Approximate Distribution}. The modips.SBS procedure sanitizes sufficient statistics in a Bayesian model, the GS of which can be challenging to obtain analytically sometimes. For an alternative, we may sanitize an approximation $g(\bs{\theta}|\x)$ to $f(\bs{\theta}|\x)$, the sanitization of which is easier.  Variational inference (VI) is a state-of-art framework for analytical approximation of $f(\bs{\theta}|\x)$ \citep{blei2017variational}; the recent normalizing flow VI approach \citep{rezende2015variational} can also yield approximate posterior samples.  There exists work that integrates DP in VI \citep{jalko2016differentially, waites2020differentially}, a relatively new area of study. 
What we propose below are three simple numerical approaches that do not involve VI but require some discretization of $f(\bs{\theta}|\x)$. The discretization incurs some information loss, which, supposedly, also brings in some some privacy protection. At the moment, we do not take this into account but rely on explicit randomization mechanisms to achieve DP. The main reason is that the  privacy guarantees associated with discretization need to be quantified (before being incorporated), an interesting topic for future research.  

The first procedure is Sanitization of Discretized Density function (SDD) (Algorithm \ref{alg:SDD}). The only step  that costs privacy is the bin selection in the discretized distribution via the exponential mechanism (line 2). Note that though $f(\bs\theta|\x)$ is listed as the input in Algorithm \ref{alg:SDD}, it does not need to be normalized and can be proportional to $f(\bs\theta|\x)$.   Cutting  $f(\bs\theta|\x)$ into $b$ bins in a $p$-dimensional space for $p\!>\!1$ can be challenging. A simpler (though likely less optimal) approach is to discretize each dimension separately and the bins across $p$ dimensions are intersections among ``marginal'' bins from each dimension and the total number of bins is $B\!=\!\prod_{j=1}^pB_j$, where $B_j$ is the number of bins in the $j^{\text{th}}$ dimension. 

\begin{algorithm}[H]
\caption{The SDD procedure}\label{alg:SDD}
\SetAlgoLined
\SetKwInOut{Input}{input}
\SetKwInOut{Output}{output}
\Input{posterior distribution $f(\bs\theta|\x)$, number of bins $b$, privacy budget $\epsilon$.}
\Output{a privacy-preserving posterior sample $\bs\theta^*$}
discretize $f(\bs\theta|\x)$ into $b$ bins $\{\mathcal{B}_i\}_{i=1,\ldots,B}$; to preserve privacy, the choice of the cut points for the bins should be independent of the observed $f(\bs\theta|\x)$\;
select a bin via the exponential mechanism of $\epsilon$-DP:  $\Pr(\mathcal{B}_i)\!\propto\!\exp(-u_i\epsilon/(2\Delta_u))$, where $u_i\!=\!-\log\big(\int_{\bs\theta\in \mathcal{B}_i}\! f(\bs\theta|\x)d\bs\theta\big)$;
%and $A_0\triangleq\sup_{\x,\bs{\theta}}\log(f(\bs{\theta}|\x))$\;
denote the selected bin by $\mathcal{B}_k$ and the index for the marginal bin in the $j$-th dimension in bin $\mathcal{B}_k$  by $j(k)$ with end points $(c_{j,j(k)-1},c_{j,j(k)})$\;
draw a sample $\bs\theta^*$ from 
uniform$((c_{1,1(k)-1},c_{1,1(k)})\!\times\!\cdots\!\times\!(c_{p,p(k)-1},c_{p,p(k)}))$.
\end{algorithm}

The integral in the utility function $u_i=\int_{\bs\theta\in \mathcal{B}_i}f(\bs\theta|\x)d\bs\theta$ of the exponential mechanism can be evaluated via numerical approaches (e.g. MC approaches). A computationally efficient approach is motivated by the mean value theorem. Specifically, we set $\int_{\bs\theta\in \mathcal{B}_i} f(\bs\theta|\x)d\bs\theta=\bar{f}_i(\bs\theta|\x)V_i$, where $V_i$ is the volume of the hyper-cube defined by the cut points $((c_{1,1(i)-1},c_{1,1(i)})\!\times\!\cdots\!\times\!(c_{p,p(i)-1},c_{p,p(i)}))$ surrounding $\mathcal{B}_i$ and $\bar{f}_i(\bs\theta|\x)$ is the average of the density values evaluated at a relatively large number of evenly distributed $\bs\theta$ points within $\mathcal{B}_i$ or can be simply set at $f((c_{1,1(i)-1}+c_{1,1(i)})/2,\cdots,(c_{p,p(i)-1}+c_{p,p(i)})/2|\x)$. Though the latter may lead to some accuracy loss, since the goal is to define and quantify a reasonable utility function $u$ for the exponential mechanism to select a bin in a private manner instead of  precisely estimating $\int_{\bs\theta\in \mathcal{B}_i}f(\bs\theta|\x)d\bs\theta$, the rough estimate would not cause material harm. With $u_i=\bar{f}_i(\bs\theta|\x)V_i$, we can calculate the sensitivity  $\Delta_u$of $u$  using Proposition \ref{pro:Deltau} (see the Appendix for the proof).
\vspace{-3pt}\begin{pro}\label{pro:Deltau}
Let $A\!\triangleq\!\sup_{\x,\bs{\theta}}|\log(f(\bs{\theta}|\x))|$. %Let $\sup_{d(\x,\x')=1,\bs\theta}|\log(f(\x|\bs\theta))-\log(f(\x'|\bs\theta))|= A$,
$\Delta_u\!=\!\max_{\mathcal{B},\bs\theta,d(\x,\x')=1}|u_i(\x)-u(\x')|$ for the utility function $u$ is $2A$. \vspace{-9pt}% If  $V_i$ is constant across $i=1,\ldots,B$, then $\Delta_u=2AV$ and the exponential mechanism is reduced to $\Pr(\mathcal{B}_i)\propto \exp(\epsilon(\log(\bar{f}_i(\bs\theta|\x))-A_0)/(2A))$.
\end{pro}
The SDD procedure can also be implemented in an MC manner by sanitizing a histogram constructed from a set of posterior samples from $f(\bs\theta|\x)$. The steps are presented in Algorithm \ref{alg:seapod}. The GS $\Delta_i$ for $\log(N^*_i/N)$ in line 2 of Algorithm \ref{alg:SDD.mc} is given in Proposition \ref{pro:GS1}; the proof can be found in the Appendix.
\vspace{-6pt}\begin{pro}\label{pro:GS1}
Let $A\!\triangleq\!\sup_{\x,\bs{\theta}}|\log(f(\bs{\theta}|\x))|$. Denote the total number of bins in the histogram  constructed from the $N$ posterior samples of $\bs\theta$ from $f(\bs\theta|\x)$ by $b$ and the count in bin $\mathcal{B}_i$ by $N_i$ for $i=1,\ldots,B$.  $\Delta_i$ for $\log(N^*_i/N)$ is $2A$. \vspace{-9pt} 
\end{pro}\vspace{-3pt}

\begin{algorithm}[H]
\caption{The SDD.MC procedure}\label{alg:SDD.mc}
\SetAlgoLined
\SetKwInOut{Input}{input}
\SetKwInOut{Output}{output}
\Input{$N$ samples of $\bs\theta$ from $f(\bs\theta|\x)$, privacy budget $\epsilon$.}
\Output{a privacy-preserving posterior sample $\bs\theta^*$.}
construct a histogram estimator of $f(\bs\theta|\x)$ given the $N$ samples of $\bs\theta$; denote the number of histogram bins by $b$\;
obtain bin count $N_i$ for $i=1,\ldots,B$ and sanitize $\log(N_i/N)$ via the Laplace mechanism: $\log(N^*_i/N)\sim\!\mbox{Lap}(\log(N_i/N),\Delta_i/\epsilon)$  to obtain a sanitized histogram\;
normalize $N_i^*$ for $i=1,\ldots,B$ so that $\sum_{i=1}^B N^*_i=1$\;
draw a sample $\bs\theta^*$ from the histogram with sanitized counts $N^*_i$ for $i=1,\ldots,B$.
\end{algorithm}

Note that though there exist approaches to obtaining privacy-preserving density estimator given a set of samples \citep{wasserman2010statistical, hall2013differential}, they are not applicable in this setting as they deal with the case where  the samples which the density estimate is formed of and is released are also the data for privacy protection. In our case, the original data $\x$ is subject to privacy protection but its density estimate of $\x$ is not the target for sanitization or release, but rather $f(\bs\theta|\x)$; in other words, we aim to limit the privacy loss of $\x$ caused by releasing samples of $\bs\theta$ from $f(\bs\theta|\x)$.  

SDD.MC is asymptotically equivalent to SDD when $N$ is large and the bin cut points are the same between the two. If $N$ is relatively small, the constructed histogram in SDD.MC  may deviate significantly from the discretized $f(\bs\theta|\x)$ in SDD, leading to loss of accuracy and supposedly some privacy protection. We present the Sanitization of Posterior Histogram Counts (SPHC) procedure in Algorithm \ref{alg:seapod} that honors the fact that $N$ has an effect on the accuracy of the histogram and privacy guarantees by sanitizing $N_i$ instead of $\log(N_i/N)$. The GS $\Delta_1$ in line 2 of Algorithm \ref{alg:SDD.mc} is given in Proposition \ref{pro:GS2}; the proof is given in the Appendix.\vspace{-3pt}
\begin{pro}\label{pro:GS2}
Let $G\triangleq\sup_{\x,\bs{\theta}}f(\bs{\theta}|\x)$. Let $b$ denote the total number of bins in the histogram  constructed from the $N$ posterior samples from $f(\bs\theta|\x)$ and $B_j$ be the number of bins in the marginal histogram in the $j$-th dimension for $j=1\ldots,p$ with the width of the $B_j$ bins by $\mathbf{h}_j=(h_{1,j},\ldots,h_{B_j,j})$. The $l_1$ GS for the bin count $N_i$ in bin $\mathcal{B}_i$ $i=1,\ldots,B$ is $\Delta_i=NG\prod_{j=1}^p h_{j(i),j}$, where $j(i)=1,\ldots,B_j$  is the index of the marginal bin in the $j$-th dimension for the $i$-th bin of the $p$-dimensional histogram.
\end{pro}\vspace{-9pt}

%(SPH)
\begin{algorithm}[H]
\caption{The SPHC procedure} \label{alg:seapod}
\SetAlgoLined
\SetKwInOut{Input}{input}
\SetKwInOut{Output}{output}
\Input{$N$ samples of $\bs\theta$ from  $f(\bs\theta|\x)$, privacy budget $\epsilon$.}
\Output{a privacy-preserving posterior sample $\bs\theta^*$.}
construct a histogram estimator of $f(\bs\theta|\x)$ with the $N$ samples of $\bs\theta$\;
obtain bin count $N_i$ for $i=1,\ldots,B$ and sanitize $N_i$ via the Laplace mechanism of $\epsilon$-DP: $N^*_i\sim\!\mbox{Lap}(N_i,\Delta_i/\epsilon)$  to obtain a sanitized histogram\;
draw a random sample $\bs\theta^*$ from the sanitized histogram with counts $N_i^*$.
\end{algorithm}

Proposition \ref{pro:GS2} suggests the GS of $N_i$ increases linearly with the number of posterior samples $N$ for a fixed $V_i=\prod_{j=1}^{p}h_{j(i),j}$. The choice for $\mathbf{h}_j$ should be independent of the local data $\x$ to not cost privacy.  Bin number determination rules can be used to calculate $\mathbf{h}_j$. For example, if the Sturge's rule is used to determine the number of bins separately for each dimension, then $B_j\equiv B=[\log_2N]+1$ and every element in $\mathbf{h}_j$ is the same  and equals $R_j/B$, where $R_j$ is the support range of the $j$-th dimension; thus $\Delta_i\!\equiv\!NG \prod_{j=1}^p(R_j/B)\!=\!NG\big(\prod_{j=1}^pR_j\big)/(\log_2N\!+\!1)^p$. Given the relationship, we can back calculate $N$ to achieve a desirable value of $\Delta_i$ for a fixed $B$. For example, if $p=1$, $\Delta_i=NGR/(\log_2N+1)$ increases with $N$ at a sub-linear rate. Say $G=0.1$ and $R=6$, setting $B=20$ (thus $h=R/B=3/10$) and $\Delta_1=1$ leads to  $N=1/(Gh)\approx24$.

The SDD, SDD.MC, and SPHC procedures can be useful even when $f(\bs\theta|\x)$ is a common distribution such as multivariate normal (MVN) distributions. Though theoretically we can sanitize the mean and covariance matrix of an MVN to achieve DP, they are functions of $\x$ and their GS can be difficult to derive depending on the Bayesian model. When applying the SDD or the SDD.MC to sanitize an MVN, we  only need to know the bounds $A$ and $G$ for $|\log(f(\bs{\theta}|\x))|$ and $f(\bs{\theta}|\x)$, respectively. The volume around a specific region in an MVN distribution does not change if the MNV is standardized (mean 0 and marginal variance in each dimension is 1) as long as the cut points around the region is relocated and scaled simultaneously. Denote the standardized MVN as  $\mathcal{N}(\mathbf{0},\mathbf{r})$,  where $\mathbf{r}$ is the  correlation matrix.  With the standardization on $\bs\theta$, we can use the same bounds $[-C,C]$ on $\bs\theta$ in each of the $p$ dimensions, where $C>0$ is large enough that beyond $[-C,C]$ there is ignorable probability mass. F Set $B_j$ the same in all $p$ dimensions and let $h\!=\!2C/B$, then the cut points for the bins in each dimension is $\mathbf{C}\!=\![-C,h\!-\!C,\cdots,C\!-\!h,C]$. %or $\{kh\!-\!C\}_{k=0,\ldots,B}$. 
Denote the left cut point in the $j$-th dimension for bin $\mathcal{B}_i$ by
$\mathbf{C}[j(i)]$, where $j(i)=1,\ldots,B+1$. For the  SPHC procedure, per Proposition \ref{pro:GS2}, $\Delta_i\!=\!NGV_i\!=\!N \max_i\{\Phi((\mathbf{C}[1(i)\!+\!1],\cdots,\mathbf{C}[p(i)]\!+\!1);\mathbf{0},\mathbf{r})\!-\!
\Phi((\mathbf{c}[1(i)],\cdots,\mathbf{C}[p(i)]);\mathbf{0},\mathbf{r})\}= N\!\left(\Phi(\mathbf{h}/2;\mathbf{0},\mathbf{r})\!-\!
\Phi(-\mathbf{h}/2;\mathbf{0},\mathbf{r})\right)$, where $\mathbf{h}_{p\times1}=(h,\ldots,h)^T$ and $\Phi$ is the CDF. The posterior correlation matrix $\mathbf{r}$ is a function of $\x$ and needs to be sanitized or specified independently of $\x$ using prior knowledge to save privacy cost. For example, since the larger the elements in $\mathbf{r}$ are, the larger $\Delta$ is, we may set all correlations in $\mathbf{r}$ at the same value that is deemed rarely large in practice to be conservative. 
This approach is employed in \cite{liu2020differentially} to obtain privacy-preserving posterior samples of ERGM parameters for social network data. 

\vspace{-3pt}\subsection{Sanitization of Statistics in modips.SBS}\label{sec:bigdata}\vspace{-9pt}
For the modips.SBS procedure, the sufficient statistics $\s$ associated with a Bayesian  model is often multi-dimensional. For a fixed privacy budget, it would be in the best interest of data users to preserve as much original information as possible when sanitizing $\s$. Toward that end, we may first examine whether the elements in $\s$ can be grouped based on the data they are calculated and leverage the parallel composition principle. Specifically, statistics in the same group share at least one individual whereas statistics in different groups are calculated based on disjoint subsets of individuals; each different group receives the full budget per the parallel composition. When it comes to sanitizing the statistics from the same group, there are different schemes for budget allocation and we introduce two below - communal sanitization and individualized sanitization. For easy illustration, we present the two definitions in the context of the Laplace mechanism; the definitions are general and apply to other mechanisms such as the Gaussian mechanisms and exponential mechanisms.  
\vspace{-3pt}\begin{defn}\label{def:con} Denote the $l_1$-GS of a multidimensional $\s$ by $\delta_{\s}=\sum_{i=1}^r\Delta_i$, where $\Delta_i$ is the $l_1$-GS of $s_i$ for $i=1,\ldots,r$. For \emph{communal sanitization}, $s_i$ is sanitized via $s_i^*=s_i +$ Laplace$(0, \delta_{\s}\epsilon^{-1})$.\vspace{-3pt}
\end{defn}
\begin{defn}\label{def:ind}  Denote the $l_1$-GS of $s_i$ in a multidimensional $\s$ by $\delta_i$ for $i=1,\ldots,r$ and by $w_i$ the proportion of $\epsilon$ allocated to $s_i$, where $\sum_{i=1}^r w_i=1$. For \emph{individualized sanitization},  $s_i$ is sanitized as in $s_i*=s_i +$ Laplace$(0, \delta_i(w_i\epsilon)^{-1})$.\vspace{-6pt}
\end{defn}
\vspace{-3pt}In short, all elements in $\s$ are sanitized via the same Laplace mechanism in the communal sanitization,  while the sanitation is ``individualized'' for each element in the individualized sanitization. Remarks \ref{rem:w} compares the communal sanitization  and individualized sanitization in two special scenarios. The proof can be found in Appendix \ref{app:conjind}.

\vspace{-3pt}\begin{rem}\label{rem:w}
(a) The communal sanitization for the Laplace mechanism is a special case of the individualized sanitization when $w_i=\delta_i\left(\sum_{i=1}^r \delta_i \right)^{-1}\!=\delta_i\delta^{-1}_{\s}\propto\delta_i$ for $i=1,\ldots,r$ in the latter.  (b) Set $w_i\equiv 1/r$ (equal allocation) in the individualized sanitization. Define the average sensitivity $\bar{\delta}_{\s}\triangleq\delta_{\s}/r$. If $\delta_i<\bar{\delta}_{\s}$, the scale parameter of the Laplace mechanism for the individualized sanitization is smaller that in the communal sanitization; in other words, if the sensitivity of an element $s_i$ in $\s$ is smaller than the average, then  allocating the same privacy budget to  $s_i$ as to every other statistic in $\s$ in the individualized sanitization leads to less perturbation compared to when the communal sanitization is employed. If the sensitivity of an element $s_i$ in $\s$ is smaller than the average, then allocating the same privacy budget to $s_i$ as to every other statistic in $\s$ in the individualized sanitization leads to more perturbation compared to when the communal sanitization is employed.
\end{rem}

\vspace{-6pt} The individualized sanitization offers more flexibility as it allows users to specify the privacy budget each $s_i$ receives.   There is no restriction on how to specify $w_i>0$ as along as $\sum_{i=1}^r w_i=1$ is satisfied. Equal allocation as given in Part (b) of Remark \ref{rem:w} may be used; one may define $w_i$ according to how  ``important'' $s_i$ is by some importance metrics.\footnote{The definition of ``importance'' varies from case to case (e.g., statistically v.s. practically important); careful considerations are required  when choosing $w_i$ according to importance.} For example, in the context of modips.SBS, if an element in $\s$ is deemed more influential in the quality of synthetic data and we can deem it  important and allocate it a big portion of $\epsilon$. 

\vspace{-3pt}\subsection{Model-free dips}\label{sec:modelfree}
\begin{figure}[!htb]\begin{center} \vspace{-9pt}
\includegraphics[scale=0.8]{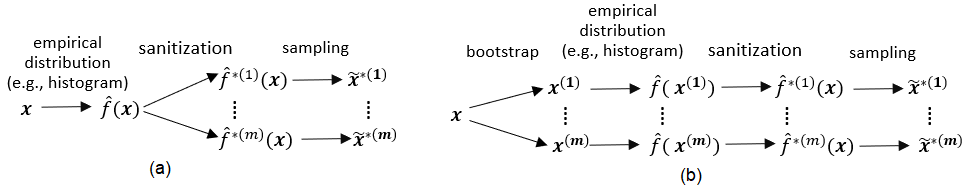} 
\caption{Examples of model-free dips schemes. (a) sampling from sanitized empirical distribution of $\x$; (b) sampling from sanitized empirical distribution of bootstrapped samples of $\x$.}\label{fig:modelfree}
\end{center}\vspace{-12pt}\end{figure}

\vspace{-9pt}
Implementation of the modips procedure requires specification of a Bayesian model given data $\x$ and sanitization  of the posterior distribution. If the model does not represent the underlying unknown population distributing $f(\x)$ well, the synthetic data can deviate significantly from the original data, leading to subsequent invalid inference of population parameters based on the synthetic data.  To circumvent this potential problem, we propose a model-free version for modips, as illustrated in  Fig \ref{fig:modelfree}(a).

First, a  empirical distribution $\hat{f}(\x)$ such as histograms is constructed from $\x$ and is then sanitized in a differentially private manner to obtain $f^*(\x)$, from which $\tilde{\x}^*$ is sampled. The synthetic data $\tilde{\x}^*$ resembles the original $\x$ excepts for the variability due to sanitization and the error in constructing $\hat{f}(\x)$ and ignores the uncertainty of the underlying unknown population distribution $f(\x)$ that the sample data $\x$ come from. This may lead to  underestimated variance for inference based on $\tilde{\x}^*$. One solution to this problem is to incorporate a bootstrap step to propagate the uncertainty from not knowing $f(\x)$ in the synthetic data, as demonstrated in Fig \ref{fig:modelfree}(b).

\vspace{-12pt}\section{Statistical Inference in  Differentially Private Synthetic Data Analysis} \label{sec:inference}\vspace{-6pt}
In this section, we  provide a framework to obtain statistical inference from multiple differentially private synthetic (dips) datasets. There exists  work in the literature on the inference in the analysis of dips data, but its often focuses on a specific type of analysis or a specific type of data.  \citet{estimation} proposed a differentially private estimator via the ``subsample-and-aggregate'' technique with a differentially private $\alpha$-Winsorized mean over the subsamples. The private estimator applies to a large class of original estimators and approximates the original mean as long as the estimators from the subsamples are i.i.d. from an approximately Gaussian distribution with a bounded third moment, for sufficiently large $n$. \citet{Charest2010} explicitly modelled the differentially private mechanism in the Bayesian inference of synthesized univariate binary data; \citet{annals} treated the Laplace mechanism as a measurement error on the sufficient statistics of the $\beta$-model for random graphs and established the conditions for the existence of the private maximum likelihood estimator  for the degree sequence in graphs that achieves the same convergence rate as non-private estimators. \citet{jrssc} applied MCMC techniques and modelled the sanitization mechanism when analyzing synthetic social networks in the framework of exponential family random-graph models. 

In summary, to appropriately capture the variability introduce during the synthesis and sanitization process in inference based on dips data, there are at least two approaches: 1) explicitly model the sanitization and synthesis mechanisms and estimate variance either analytically or computationally (e.g., MC sampling in the Bayesian framework); 2) propagate  uncertainty through releasing multiple synthetic datasets (MS) and apply an appropriate variance combination rule.  The former does not require the release of multiple synthetic datasets but users of the synthetic data need to be provided with full details of the sanitization mechanisms so to model and incorporate the mechanism in their data analysis procedure. This can be very challenging especially for users who are not familiar with DP and sanitization mechanisms; even for users who are familiar of DP, incorporating a randomization mechanism in a commonly used analysis procedure may brings analytical and computational challenges. In contrast, the MS approach is more friendly to data users as they do not need to explicitly model the sanitization mechanism and can analyze the synthetic data as they would for the original data. The only additional step on top of what they would normally do is to repeat the analysis procedure $m$ times, one for each of the $m$ sets of synthetic data and then combine the inferences to generate the final inferential results. For this reason, we focus on the MS approach. The main results are provided in Theorem \ref{theorem:varcomb}. Before that, we first present Def \ref{def:consistency} on which  Theorem \ref{theorem:varcomb} is based. 
\vspace{-6pt}\begin{defn}[consistency of sanitized posterior  distribution]\label{def:consistency} Suppose $f^*(\bs\theta|\x)$ is a sanitized version of the posterior distribution $f(\bs\theta|\x)$  via a differentially private mechanism with privacy loss $\epsilon$. If $f_{\epsilon}^*(\bs\theta|\x)\overset{d}{\rightarrow} f(\bs\theta|\x)$ as $\epsilon\rightarrow\infty$, then $f_{\epsilon}^*(\bs\theta|\x)$ is consistent for $f(\bs\theta|\x)$.
\end{defn}
\vspace{-12pt}\begin{thm}  \label{theorem:varcomb}
Assume the model from which the posterior distribution of  $\bs\theta$ is obtained in the modips procedure is the same as the one used for synthesis and that $f^*(\bs\theta|\x)$ is consistent for $f(\bs\theta|\x)$. Let  $\tilde{\x}^{*(i)}$ denote the $i$-th synthetic dataset for $i=1,\ldots,m$ given a $\bs\theta^{*(i)}$ sample from the differentially private posterior distribution $f^*(\bs\theta|\x)$. Denote  the parameter of inferential interest by $\beta$ and assume the statistical procedure for obtaining inference for $\beta$ is the same given $\tilde{\x}^{*(i)}$ or $\x$. Denote the estimate of $\beta$ from $\x$ and the corresponding variance estimate  by $\hat{\beta}$ and $v$, those based on $\tilde{\x}^{*(i)}$ by $\hat{\beta}^{*(i)}$ and $\hat{v}^{*(i)}$, respectively. If $\hat{\beta}\xrightarrow[]{p}\beta, \hat{\beta}^{*(i)}\xrightarrow[]{p}\beta^*,  \E\big(\hat{\beta}^{*(i)}|\x\big)\rightarrow \hat{\beta}, \E\big(m^{-1}\sum_{i=1}^m\mbox{V}(\beta^{*(i)}|\tilde{\x}^{*(i)})|\x\big)\!\rightarrow\!\mbox{V}(\beta|\x)$, and 
$\E\big((m\!-\!1)^{-1}\sum_{i=1}^m  (\hat{\beta}^{*(i)}\!-\!\bar{\beta}^*)^2|\x\big)\!\rightarrow\! \mbox{V}(\beta^{*(i)}|\x)$, where $\bar{\beta}^*=\textstyle m^{-1}\sum_{k=1}^m\hat{\beta}^{*(i)}$, as $n\rightarrow\infty$, then\vspace{-9pt}
\begin{itemize}[leftmargin=16pt,itemsep=0pt]
\item[a)] $\bar{\beta}^*=\textstyle m^{-1}\sum_{k=1}^m\hat{\beta}^{*(i)}$ is a consistent estimator for $\beta$;
\item[b)] an asymptotically unbiased estimator for the variance of $\bar{\beta}^*$ is\vspace{3pt}
\begin{align*}
u=\varpi+m^{-1}b,&\mbox{ where } \varpi=\textstyle m^{-1}\sum_{k=1}^m\hat{v}^{*(i)} \mbox{ and }  b=\textstyle m^{-1} \sum_{i=1}^m (\hat{\beta}^{*(i)}-\bar{\beta}^*)^2.%\label{eqn:var}
\end{align*}

$\varpi$ is the averaged within-set variance  $\E(\V(\theta|\x)|\x^{*(1)},\ldots,\x^{*(m)})$. $b$ is the between-set variance  $\V(\E(\theta|\x)|\x^{*(1)},\ldots,\x^{*(m)})$ that comprises two components $b_1$ and $b_2$; $b_1$ is the variability incurred by sanitization and $b_2$ is the variability due to synthesis that further comprises $b_{21}$ and $b_{22}$, corresponding to the posterior variability of $\theta$ and the sampling variability of $\x$ given $\theta$, respectively;
\item [c)] the inference of $\beta$ given $\tilde{\x}^{*(1)},\ldots, \tilde{\x}^{*(m)}$ is based on $t_{\nu}(\bar{\beta}^*, m^{-1}b+\varpi)$, where the degree of freedom $\nu =(m-1)(1+m\varpi/b)^2$.
\end{itemize}\vspace{-15pt}
\end{thm}
The proof is provided in Appendix \ref{app:varcomb}. Though Part b) decomposes the between-set variance $b$ component, there is seldom an interest in  quantifying  $b_1$ and $b_2$ separately. If there is such a need, it can be fulfilled via the nested modips procedure depicted in Fig \ref{fig:modips3}.

We expect the results in Theorem \ref{theorem:varcomb} also apply to general dips approaches including  both model-based and model-free dips approaches that do not generate synthetic data from differentially private posterior predictive distributions. Depending on the procedure used, the nature of between-set variability $b$ may be different. For example, if we use a model-free dips approach as given in Fig \ref{fig:modelfree}(a), then $b$ comprises the sanitization variability $b_1$ and part of the sanitization variability $b_2$ (i.e., the sampling variability $b_{22}$ from $\hat{f}(\x)$  but not the  uncertainty due to not knowing the distribution of $\x$). Note that this is not the problem of the variance  formula $\varpi+b/m$, but rather because the dips approach in Fig \ref{fig:modelfree}(a) does not take into account the fact that the population distribution $f(\x)$ is unknown. If the procedure in Fig \ref{fig:modelfree}(b) is used, then  $b$ includes $b_1$ and both components of $b_2$ as the uncertainty of $f(\x)$ is captured through the bootstrap step.

The formula  $u=\varpi+m^{-1}b$  in Theorem \ref{theorem:varcomb} coincides with that for combining inferences from MS data generated by the partial sample synthesis in the non-DP setting \citep{reiter2003} that is also applicable to the non-DP full sample synthesis.\footnote{ Full sample synthesis can be viewed as a special case of partial sample synthesis with a 100\% synthesis proportion.}  Further analysis suggests that the equivalence between our formula developed for modips and that for the partial/full sample synthesis in the non-DP setting is not coincidental as the modips procedure can be viewed as a differentially private version of the full sample synthesis with an extra step  of explicitly sanitizing the posterior distribution; or we can regard the full sample synthesis as the asymptotic case of  modips approach as $\epsilon\rightarrow\infty$.  

In the traditional non-DP DS setting, the choice of $m$ is mostly driven by computational time and storage considerations; thus small $m$  is preferred as long as it is large enough to capture the between-set variance and deliver valid inference. The empirical studies \citep{reiter2003, raghunathan2003multiple} suggest small $m$ (e.g., $\le 10\sim15$) seems to work. In contrast, in the DP setting, the decision on $m$ is driven by the utility of the sanitized data at a pre-specified privacy budget $\epsilon$ and it is not necessarily true that a larger $m$ yields better utility in the synthetic datasets  overall. This is each synthesis receives only $1/m$ of the total budget. While a too small $m$ may not be sufficient to capture the $b$ component well, a too large $m$ risks spreading $\epsilon$ too thin over $m$ sets and each synthetic set is so``over-perturbed'' that aggregating information across $m$ synthetic set cannot remedy the information loss.  

We examine the effect of $m$ on the inference in dips data empirically in Sec \ref{sec:examples} and plan to provide more  theoretical analysis on this problem in the future, which can be a challenging task. Our expectation is that $m$ synthetic datasets differ more and more as $m$ increases, leading to increase in $b$ at least initially. On the other hand, if $m$ is very large, the large amount of perturbation may push each synthetic data to some consistent extremes, causing $b$ to decrease instead. In addition, $\varpi$ may also change with $m$ in a manner that depends on how much the randomization mechanism perturbs $\x$ in what way. If $m$ does affect $\varpi$,  the effect is expected to be smaller compared to that on $b$ (see the experimental results in Sec \ref{sec:var}). The difficulty in the theoretical analysis lies in  obtaining a functional form for $u(m)=\varpi(m)+m^{-1}b(m)$, which may vary case by case. If  close form for $u(m)$ exists for some problems,  the rate of the change of $u(m)$ with $m$ is quantified by its first derivative  $\varpi'(m)+m^{-1}b'(m)-m^{-2}b(m)$. If $m^2\varpi'(m)+mb'(m)<b(m)$, then $u(m)$ decreases with $m$; otherwise, $u(m)$  increases with $m$. Whether there exists an $m$ the leads to a minimum or maximum in $u$ depends on the specific problem (see the experimental results in Sec \ref{sec:m}) and meaningful general theoretical results may not exist.  

%%%%%%%%%%%%%%%%%%%%%%%%%%%%%%%%%%%%%%%%%%%%%%%%%%%%%%%%%%%%%%%%%%%%%%%%%%%

\vspace{-3pt}\section{Numerical Examples} \label{sec:examples}\vspace{-9pt}
This section present numerical examples to (1)  examine the effects of $m$ on the inference in the synthetic data;  (2) demonstrate the validity of the inferential procedure in Theorem \ref{theorem:varcomb}; (3) investigate the impact of the budget allocation on the inference in the synthetic data for the modips.SBS procedure. It also surveys and summarizes the results from the published work that has implemented modips procedure or the inferential rule in Theorem \ref{theorem:varcomb}.

\vspace{-6pt}\subsection{Impact of $m$ on Inference}\label{sec:m}\vspace{-6pt}
We present two examples. The first example is  binary data and the parameter of inferential interest is proportion $p$; the second example is Gaussian data and the parameter of interest is mean $\mu$  ($\sigma^2$ is assumed known and set at 1). For the binary case, we examine two values on $p$ (0.1, 0.5; $p=0.1$ represents an unbalanced data scenario). For the Gaussian case, we set $\mu=0$,  WLOG. We applied the modips.SBS approach and used the Laplace mechanism to sanitize sufficient statistics.  With the uniform priors for $p$ and $\mu$, the sufficient statistic  in the posterior distribution of $p$ is the sample proportion and that of $\mu$ is the sample mean. The $l_1$-GS of $p$ and $\mu$ is  $1/n$ and  $(c_1-c_0)/n$, respectively, where $[c_0,c_1]$ are the bounds for the Gaussian data \citep{Liu2019}. We examine two sets of bounds: $[c_0,c_1]=[-4,4]$ and $[4,5]$. Given $\Pr(|X|\!>\!4)\!=\!0.0063\%$, the truncated data can still be well approximated by a Gaussian distribution. The sanitized statistics via the Laplace mechanism can be out of bounds ($<0$ or $>1$ for the sanitized sample proportion, and $<c_0$ or $>c_1$ for the sanitized sample mean and the synthetic data in the Gaussian case) as the support for the Laplace distribution is the real line. There are two ways to legitimize the out-of-bound values -- truncation and boundary inflated truncation (BIT) \citep{Liu2019}. The former throws away out-of-bound values and the latter sets the values smaller than the lower bound at the lower bound and those larger than the upper bound at the upper bound. We examine $m\in[2,500]$ with the understanding that large $m$ is for investigation purposes and unlikely  used in practice. The overall privacy budget is $\epsilon=1$, equally shared across $m$ synthetic datasets. The inferences for $p$ and $\mu$ were obtained via the formulas in Theorem \ref{theorem:varcomb} and summarized over 5,000 repetitions. 

\begin{figure}[!htb]
\begin{center}  
\includegraphics[width=0.98\textwidth]{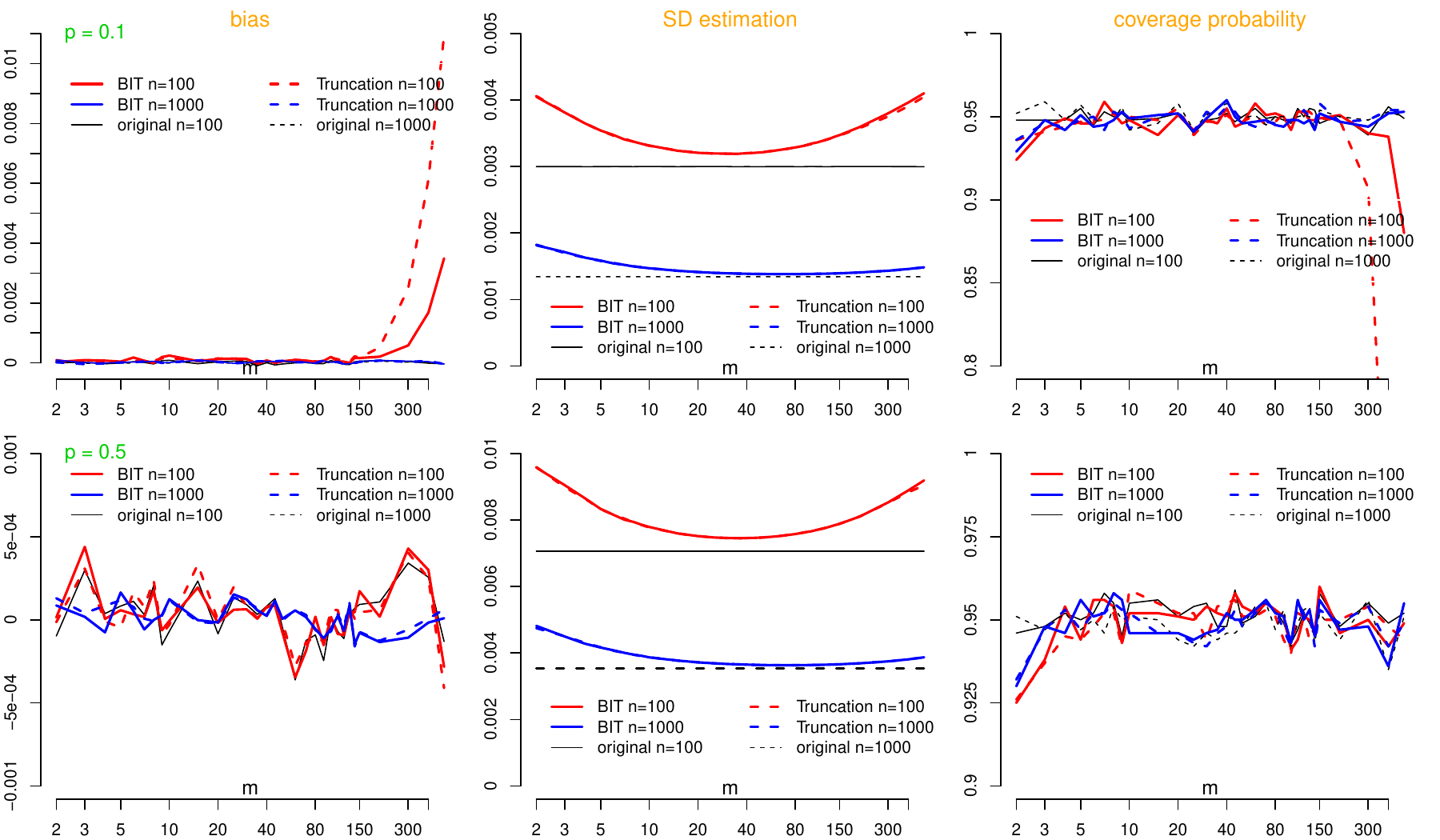} 
\footnotesize{(a) binary data}
\includegraphics[width=0.98\textwidth]{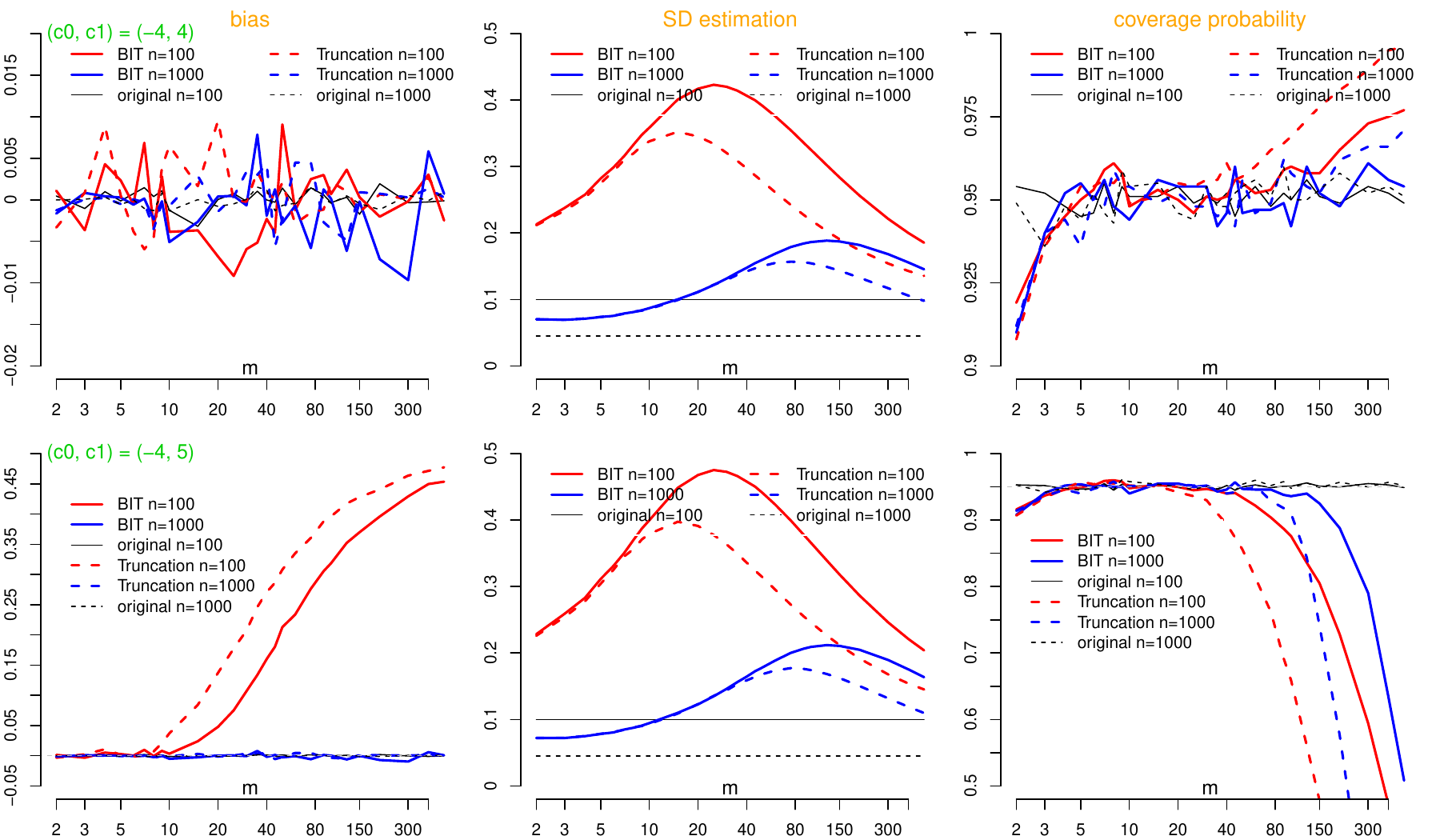} 
\footnotesize{(b) Gaussian data}
\caption{Inference based on synthetic data via modips.SBS at $\epsilon=1$ ($m$-axis is on the log scale)}\label{fig:m}
\vspace{-24pt}\end{center} \end{figure}
Fig \ref{fig:m} presents the bias, standard deviation (SD) estimation, and coverage probability (CP) of the 95\% CIs in the estimation of the parameters.
The main observations in the binary case are as follows. (1) There is minimal bias in $\bar{p}^*$ (point estimate of $p$) for all the examined $m$ in almost all the simulation scenarios except for the slight over-estimation (bias $\sim0.01$) for $m>200$ when $n=100$  in the case of $p=0.1$; such a large $m$ is unlikely to employ in practical applications. (2) The SD estimate for $\bar{p}^*$ based on the sanitized data is larger than that based on the original, which is expected given the sanitization and synthesis. The SD first decreases as $m$ increases, gets close to the original SD for $n=1,000$, and remains roughly constant for $m>30\sim40$. When $n=100$, the variance also first decreases with $m$, hits its minimum around $m=30\sim40$, and then increases with $m$. (3) The CP is near nominal (95\%) for almost all $m$ with the slight under-coverage but still $\sim92.5\%$ at $m=2$ and severe under-coverage when $m>200, n=100, p=0.1$ -- a consistent observation with the large bias in the same condition.  For the Gaussian case, $m$ has a smaller impact on the inference when the bounds are symmetric  $(-4,4)$ around the true mean ($\mu=0$) compared to the asymmetric bounds $(-4,5)$. Specifically, (1) there is minimal bias in the estimation of $\mu$  for all $m$ in the symmetric bounds case and obvious bias for $m>10$ in the asymmetric bounds case when $n=100$. (2) How the variance estimate changes with $m$ is similar between the symmetric and asymmetric bounds, though the estimate is slightly smaller in the former. When $n=1,000$, the variance estimate remains roughly constant for $m<10$, then increases with $m$, reaches its peak around $m=50$ (truncation) to $150$ (BIT), and then decreases with $m$. When $n=100$, the variance increase with $m$, reaches its peak around $m=20$ (truncation) to $m=40$ (BIT), and then decreases with $m$. (3) The CP is near nominal (95\%) across most $m$ with slight under-coverage ($\sim91\%$) at $m=2$ and some over-coverage for $m>40$  when $n=100$ when the bounds are symmetric. For the asymmetric bounds case, there is severe under-coverage for $m>20$ (truncation) and $m>40$ (BIT) when $n=100$, and for  $m>80\sim150$ when $n=1,000$. 

In summary, the observations in Fig \ref{fig:m} suggest that $m$ affects the quality of inference based on sanitized data and how it affects  depends on the true parameter value, the global bounds on the data, the truncation schemes, etc. Among bias, variance estimate, and CP, the variance estimate is consistently the most sensitive to the change in $m$ value. Finally, at least in these two examples, $m\in[3,10]$ seems to be a good choice for satisfactory performances in bias, variance estimate, and CP; a large $m$ is unnecessary from a computation and storage perspective and sometimes undesirable from an inferential perspective. 

\vspace{-6pt}\subsection{Validation of the Variance Combination Rule} \label{sec:var}
\begin{figure}[!htb]\centering
\includegraphics[scale=0.40]{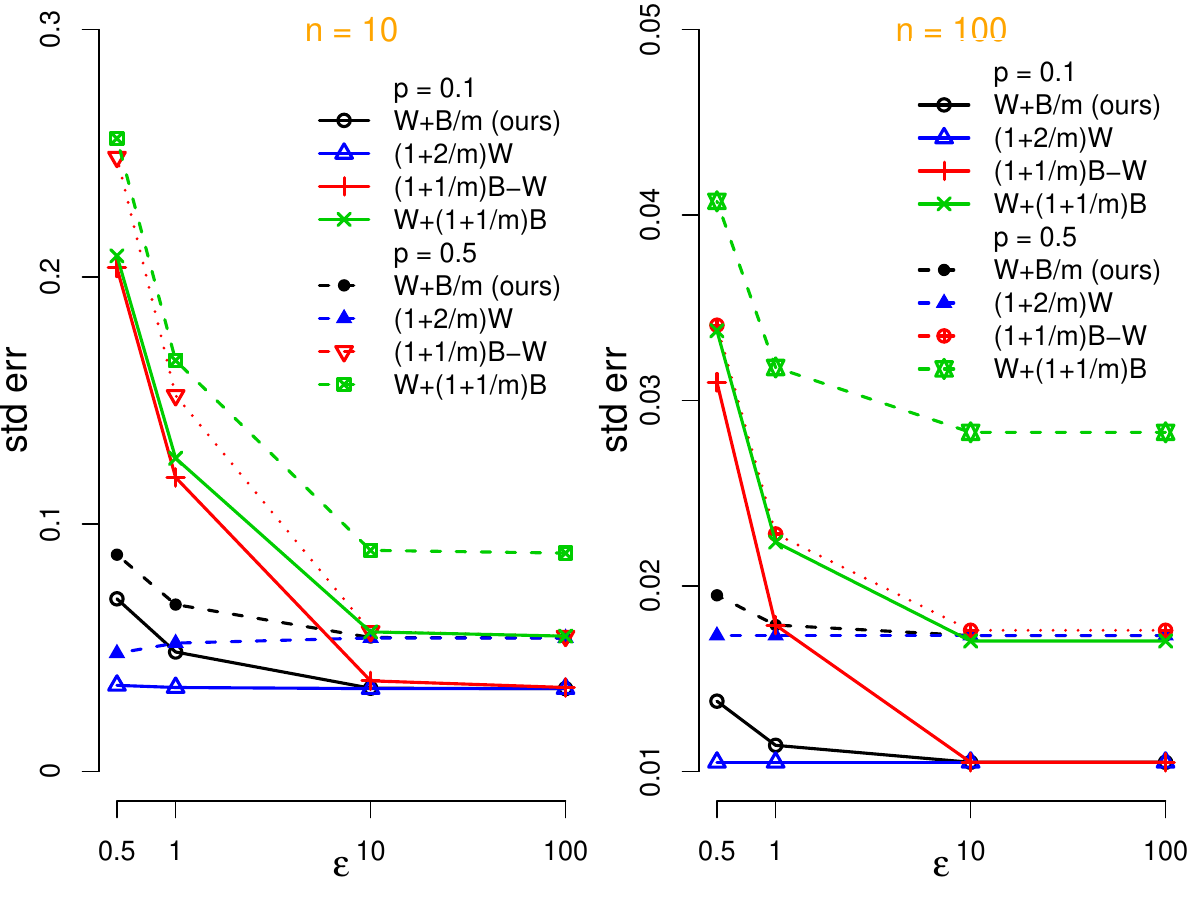}
\includegraphics[scale=0.40]{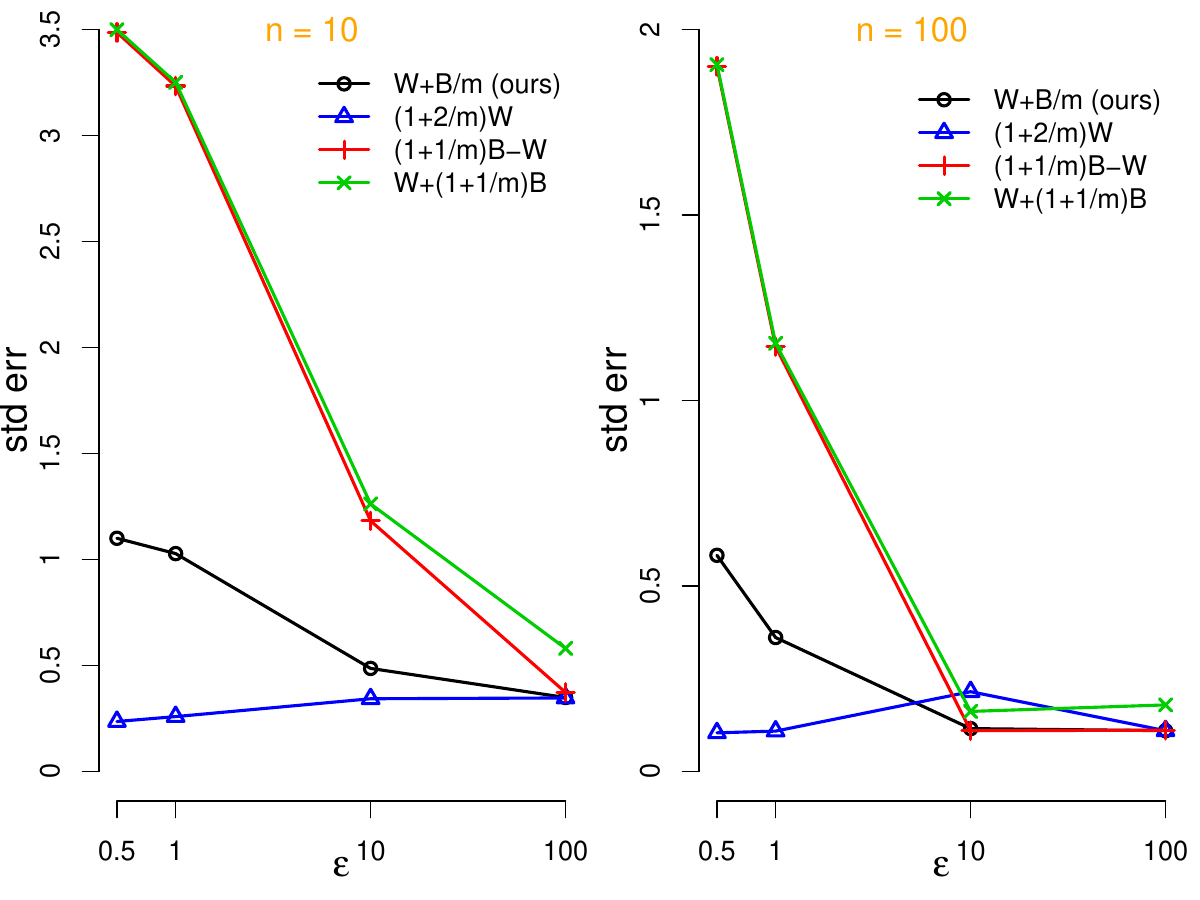}\vspace{-7pt}\\
\footnotesize{(a) binary data\hspace{5cm}(b) Gaussian data}\\
\caption{Standard error estimate via different variance combination rule} \label{fig:var}\vspace{-12pt}
\end{figure}

\vspace{-6pt}
We use similar simulation settings as in Sec \ref{sec:m} to examine and  compare the variance combination rule $\varpi+m^{-1}b$ in Theorem \ref{theorem:varcomb} with three other variance combination rules that are developed in different but related settings: (1)  $(1+m^{-1})b+\varpi$ that combines inferences from multiply imputed datasets in the setting of missing data \citep{MIbook}; (2)  $(1+ m^{-1})b-\varpi$ developed by \citet{raghunathan2003multiple} for inference in the population full synthesis in the non-DP setting; (3)  $(1+2/m)\varpi$ proposed by \citet{raab2017practical} for the multiple synthesis in non-DP setting.\footnote{As stated in Sec \ref{sec:inference}, the formula in \citep{reiter2003} for the partial sample synthesis in the non-DP setting is the same as in Theorem \ref{theorem:varcomb} and there is no need to include it as a comparison method.} We used $m=10$ in all cases.  Since the results for asymmetric and symmetric bounds $[c_0,c_1]$ in the Gaussian case are similar and  the results from the two bounding schemes (BIT and truncation) are similar,  we  present those from the symmetric bounds and the BIT scheme only (Table \ref{tab:rule} and Fig \ref{fig:var}).

The main observations on the CP from Table \ref{tab:rule} are summarized as follows. 1) Our proposed variance rule $m^{-1}b+\varpi$ provides nominal coverage in all simulation scenario. (2) $(1+2/m)b$ provides nominal coverage when $\epsilon$ is large $(\ge10)$ and $n=100$, but leads to severe under-coverage when $n$ or $\epsilon$ is small. (3) $(1+1/m)b-\varpi$ leads to either under-coverage or over-coverage and hardly produces near nominal coverage. (4) $(1+ m^{-1})b+\varpi$ is overly conservative and delivers close to 100\% coverage in all simulation scenarios. (5) As expected, the single synthesis leads to severe under-coverage as it does not capture the synthesis and sanitization uncertainty unless users explicitly model the synthesis and sanitization process. 
\begin{table}[!htb]
\begin{center}
\caption{Coverage probability of 95\% CI using different variance combination rules ($m=10$)}\label{tab:rule}
\begin{tabular}{c}
\small{(a) Binary Data}\\
\end{tabular}\\
\resizebox{1\textwidth}{!}{
\begin{tabular}{ccc |c| c|c|c|c| c}
\hline
\multicolumn{3}{c|}{scenario} & original & \multicolumn{4}{c|}{multiple synthesis}& single \\
\cline{1-3}\cline{5-8}
$\!\epsilon\!$ & $\!n\!$ & $\!p\!$ & & $\!B/m\!+\!W\!$ (Thm \ref{theorem:varcomb}) & $(1\!+\!2/m)\varpi$ & $(1\!+\!1/m)B\!-\!\varpi$ &  $(1\!+\!1/m)B\!+\!\varpi$  &synthesis\\
% & & & & (Thm \ref{theorem:varcomb}) & $+\varpi$ & $-\varpi$ & $+\varpi$ & \\
\hline
%$10^7$ & 10&  0.5  & 0.938 & 0.941 & 0.942  & 0.786 & 0.996 & 0.732\\
%$10^7$ & 10&  0.1  & 0.935 & 0.953 & 0.954  & 0.795 & 0.996 & 0.743\\
%$10^7$ & 100& 0.5  & 0.944 & 0.945 & 0.946  & 0.790 & 0.998 & 0.726\\
%$10^7$ & 100& 0.1  & 0.956 & 0.951  & 0.952 & 0.798 & 0.999 & 0.746\\
%\hline
$100$  & 10&  0.5  & 0.946 & 0.948 & 0.948 & 0.798  & 0.999 & 0.736\\
$100$  & 10&  0.1  & 0.935 & 0.950 & 0.949 & 0.793  & 0.996 & 0.738\\
$100$  & 100& 0.5  & 0.945 & 0.952 & 0.952 & 0.797  & 0.999 & 0.747\\
$100$  & 100& 0.1  & 0.949 & 0.949 & 0.950 & 0.791  & 0.998 & 0.743\\
\hline
$10$   & 10& 0.5  & 0.942 & 0.945 & 0.945 & 0.814 & 0.998  & 0.723\\
$10$   & 10& 0.1  & 0.930 & 0.946 & 0.946 & 0.835 & 0.997 & 0.732\\
$10$   & 100& 0.5 & 0.946 & 0.947 & 0.947 & 0.792 & 1.000 & 0.738\\
$10$   & 100& 0.1 & 0.952 & 0.948 & 0.950 & 0.795 & 0.998 & 0.747\\
\hline
$1$    & 10& 0.5  & 0.938 & 0.947 & 0.865 & 0.994  & 1.000 & 0.730\\
$1$    & 10& 0.1  & 0.936 & 0.961 & 0.840 & 0.994 & 0.999 & 0.726\\
$1$    & 100& 0.5 & 0.948 & 0.946 & 0.938 & 0.898 & 0.999 & 0.746\\
$1$    & 100& 0.1 & 0.950 & 0.952 & 0.931 & 0.958 & 1.000 & 0.749\\
\hline
$0.5$  & 10& 0.5  & 0.942 & 0.941 & 0.707 & 0.999 & 1.000 & 0.715\\
$0.5$  & 10& 0.1  & 0.928 & 0.946 & 0.532 & 0.998 & 0.999 & 0.686\\
$0.5$  & 100& 0.5 & 0.955 & 0.953 & 0.924 & 0.972 & 1.000 & 0.730\\
$0.5$  & 100& 0.1 & 0.949 & 0.949 & 0.869 & 0.992 & 1.000 & 0.756\\
%\hline
%$0.1$  & 10& 0.5  & 0.949 & 0.946 & 0.328 & 0.998 & 0.998 & 0.502\\
%$0.1$  & 10& 0.1  & 0.928 & 0.690 & 0.045 & 0.998 & 0.999 & 0.368\\
%$0.1$  & 100& 0.5 & 0.948 & 0.951 & 0.521 & 1.000 & 1.000 & 0.685\\
%$0.1$  & 100& 0.1 & 0.954 & 0.948 & 0.394 & 1.000 & 1.000 & 0.636\\
\hline
\end{tabular}}\\
\begin{tabular}{c}
\vspace{-12pt}\\
\footnotesize{(b) Gaussian Data}\\
\end{tabular}\\
\resizebox{0.95\textwidth}{!}{
\begin{tabular}{cc|c |c|c|c|c| c}
\hline
\multicolumn{2}{c|}{scenario} & original & \multicolumn{4}{c|}{multiple synthesis} & single \\
\cline{1-2} \cline{4-7}
$\epsilon$ & $n$ && $\!B/m\!+\!W\!$ (Thm \ref{theorem:varcomb}) & $(1\!+\!2/m)\varpi$ & $(1\!+\!1/m)B\!-\!\varpi$ &  $(1\!+\!1/m)B\!+\!\varpi$  &synthesis\\
% & & &  (Thm \ref{theorem:varcomb}) & $+\varpi$ & $-\varpi$ & $+\varpi$& \\
\hline
%$10^7$& 10 & 0.951 & 0.949 & 0.948 & 0.790 & 0.998 & 0.731 \\
%$10^7$& 100& 0.956 & 0.951 & 0.951 & 0.794 & 0.988 & 0.737\\
%\hline
$100$&  10 & 0.951 & 0.953 & 0.946 & 0.825 & 0.997 & 0.743\\
$100$& 100 & 0.949 & 0.952 & 0.952 & 0.795 & 0.997 & 0.741\\
\hline
$10$&  10  & 0.951 & 0.952 & 0.836  & 0.999 & 1.000 & 0.720 \\
$10$&  100 & 0.948 & 0.946 & 0.933 & 0.924 & 0.998  & 0.739\\
\hline
$1$&   10  & 0.948  & 0.951& 0.392 & 0.998  & 0.999  & 0.450\\
$1$&   100 & 0.953 & 0.956 & 0.454 & 1.000  & 1.000 & 0.674\\
\hline
$0.5$& 10  & 0.951 & 0.954 &0.334  & 0.997 & 0.998 & 0.275\\
$0.5$& 100 & 0.950 & 0.951 & 0.274 & 1.000 & 1.000 & 0.551\\
\hline
\end{tabular}} \end{center}\vspace{-24pt}
\end{table}

Fig \ref{fig:var} plots the actual SD estimates via the 4 difference variance combination rules. In the binary case, $(1+m^{-1})b+\varpi$ produces the largest SD estimate, as expected, followed by  $(1+ m^{-1})b-\varpi$,  our rule $\varpi+b/m$, and finally $(1+2/m)\varpi$. $\varpi+b/m$ and $(1+2/m)\varpi$ are similar when $\epsilon>1$ for all the examined simulation scenarios. When $\epsilon>1$, $(1+ m^{-1})b-\varpi$ also yields similar results to $\varpi+b/m$ and  $(1+2/m)\varpi$. For the Gaussian case,  $(1+m^{-1})b+\varpi$ and $(1+ m^{-1})b-\varpi$ produce very similar results, followed by $\varpi+b/m$ and then $(1+2/m)\varpi$; the latter two are similar for $\epsilon>10$. All four rules are similar at $\epsilon=100$ when $n=10$ and for $\epsilon>10$ when $n=100$. In both the binary and Gaussian cases, the SD estimate is roughly constant across $\epsilon$ for $(1+2/m)\varpi$ as the formula ignores $b$ which changes drastically with $m$ in the DP setting. This implies that $(1+2/m)\varpi$ is invalid in the DP setting.  For the other three rules that contain the $b$ component , the SD estimate stabilizes after a certain $\epsilon$, the value of which depends on $n$, the data type, and possibly the model, etc. 

In summary, the variance combination rule $(1+2/m)\varpi$ ignores the $b$ component, which matters in the DP setting, leading to under-estimated variance and under-coverage, and is thus invalid. $(1+1/m)b+\varpi$  is overly conservative for synthetic data in the DP setting. $(1+1/m)b-\varpi$ does not provide the correct combination between $b$ and $\varpi$ either as it mis-focuses on $b$ when $\varpi$ is the main contributor to the total variance at large $\epsilon$ and tends to over-weigh $b$ when it is large at small $\epsilon$. All taken together, none of the three is suitable for inference in the analysis of synthetic data in the DP setting.  In addition, even the variance by $(1+1/m)b-\varpi$ is no smaller than our $m^{-1}B+\varpi$, the former stills leads to under-coverage as inference in the former is based on the standard Gaussian distribution and the latter is based on a $t$-distribution, the degree of freedom of which is a function of $w,b$, and $m$. Finally, $(1+1/m)b-\varpi$ is less than $(1+1/m)b+\varpi$ by $2\varpi$; but the difference is not obvious for small $\epsilon$. This is because $b$ is the dominant contributor to the total variance at small $\epsilon$ and overshadows the contribution from $\varpi$.  The difference at large $\epsilon$ between the two is more obvious in the binary case than in the Gaussian case.

\vspace{-3pt}\subsection{Impact of Budget Allocation on Inference}\label{sec:w}
\begin{figure}[!htb]\vspace{-5pt}
\begin{center}  
\includegraphics[scale=0.54]{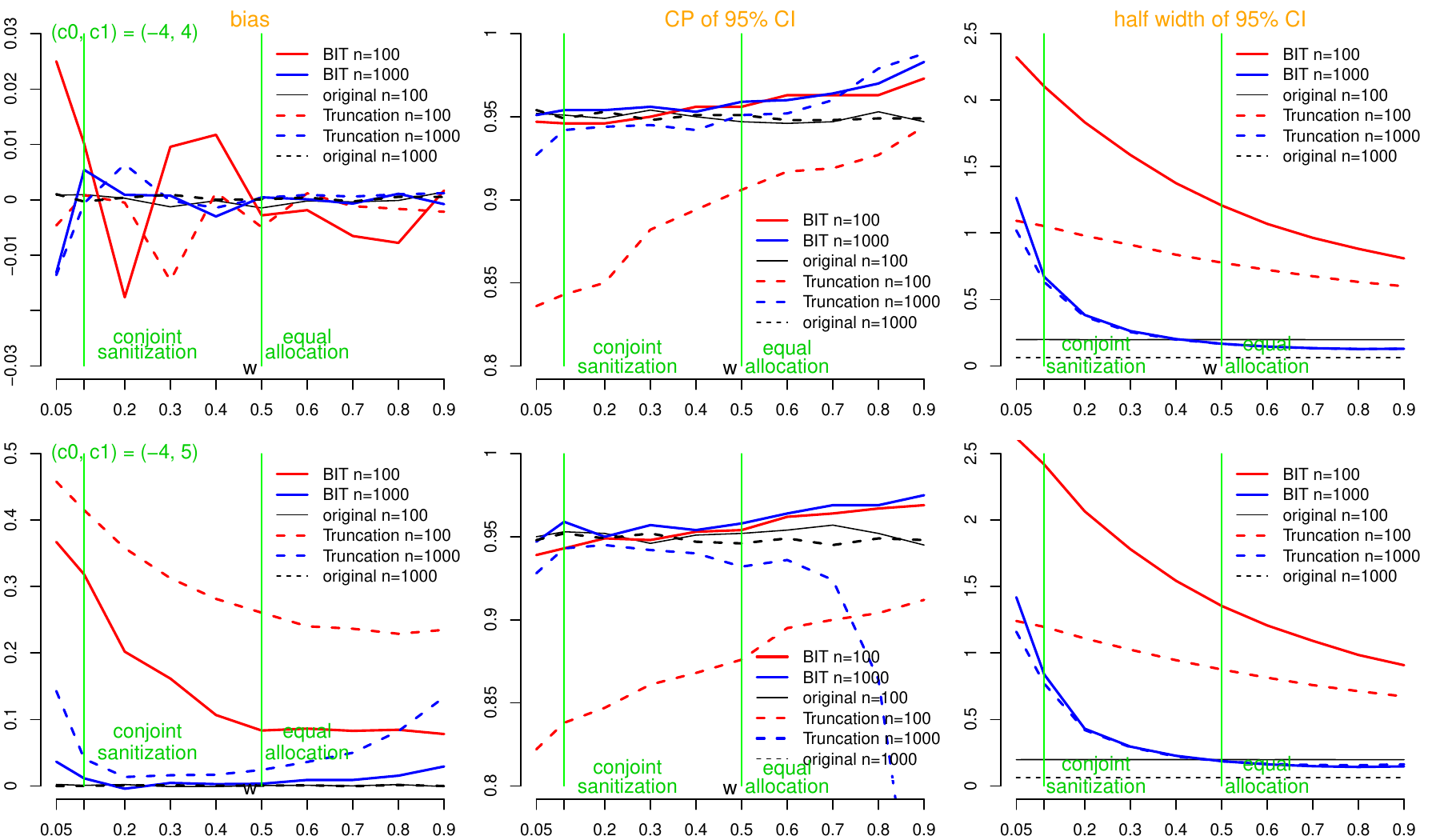}
\caption{Effect of budget allocation on inference of $\mu$ based on sanitized data ($m\!=\!10, \epsilon=1$)}\label{fig:w}
\vspace{-12pt}\end{center} \vspace{-12pt}\end{figure}

\vspace{-7pt}
This experiments examines how budget allocation schemes in the  individualized sanitizations (Definitions \ref{def:con} and \ref{def:ind})  affects the inference based on synthetic data. The data $\x$ were simulated from $\mathcal{N}(\mu=0,\sigma^2=1)$, where both $\mu$ and $\sigma^2$ are unknown. We examine both the symmetric and asymmetric bounds scenarios $(-4,4)$ and $(-4,5)$ on data as well as both the BIT and truncation bounding schemes.  The Bayesian sufficient statistics in the posterior distribution of $f(\mu,\sigma^2|\x)$ given the Gaussian likelihood and the prior $f(\sigma^2)\propto\sigma^{-2}$ is the sample mean and variance $\bar{x}$ and $s^2$, the $l_1$ GS of which is $(c_1-c_0)/n$ and $(c_1-c_0)^2/n$, respectively \citep{Liu2019}. Denote the proportion of the total privacy budget $\epsilon$ allocated to sanitizing  $\bar{x}$ and $s^2$ by $w\in(0,1)$ and $1-w$, respectively. When $w=(c_1-c_0)/((c_1-c_0)+(c_1-c_0)^2)=1/(1+c_1-c_0)$, the individualized sanitization becomes the communal sanitization. Note that the bounding scheme is not only applied to the sanitized mean and the sanitized data, which are $(\in(c_0,c_1))$, but also to the sanitized variance, which is $\in(0,(c_1-c_0)^2/4)$. We examine the effect of $w$ on the inference of $\mu$ based on the sanitized data ($m=10,\epsilon=1$). The results on the bias, the CP and the half width of the 95\% CI are presented in Fig \ref{fig:w} over 5,000 repeats. It should be noted that the inference for $\mu$ in Fig \ref{fig:w} are less accurate and precise compared to the results for the Gaussian example in Sec \ref{sec:m}. This is because the same privacy budget is used to sanitize two statistics (sample mean and variance) instead of one in this exampple.

The findings are summarized as follows. First, when the data bounds $(c_0,c_1)=(-4,4)$ are symmetric around $\mu$, $w$ barely affects the accuracy of the point estimate $\bar{\tilde{x}}^*$ for $\mu$ except for some numerical fluctuation. When the bounds are asymmetric, there is obvious bias at $n=100$, which decreases as $w$ increases and remains roughly constant for $w>0.5$. Second, the CPs are around the nominal level 95\% with some slight over-coverage for $w\ge0.5$ for the BIT bounding scheme. For the truncation bounding scheme, there is obvious under-coverage for all $w$ when $n=100$ and for $w>0.5$ when $n=1,000$. Third, it is expected that the half width of the 95\% CI based on the sanitized data is the larger than the original CI half width given the additional variability caused by sanitization and synthesis in modips. Specifically, the half width is close to the original for $w>0.5$ at $n=1,000$ and significantly deviates from the original for all $w$ at $n=100$ but decreases as $w$ increases. 

In conclusion, larger $w$ (portion of budgets allocated to sanitizing $\bar{x}$ tends to offer more precise inference for $\mu$ with non-inferior accuracy than lower $w$. The equal allocation scheme, $w=0.5$ in this case, is a reasonable and convenient choice for this example; the ``default'' communal sanitization might not lead to the most efficient or accurate inference for parameters of inferential interest. 

\vspace{-6pt}
\setstretch{1}
\subsection{Summary of results in the literature on the application of modips and inferential combination rule }\vspace{-6pt}
\setstretch{1.03}
The modips approach and Theorem \ref{theorem:varcomb} have been employed in some later work on dips in the literature citing the 2016 arxiv version of this paper. We summarize some of those results below.

\citet{RON-Gauss2019} proposed the RON-Gauss approach that combines dimensionality reduction via the random orthonormal projection and the Gaussian generative model to synthesize differentially private data. They run experiments to compare the RON-Gauss approach with 4 other dips methods, including the modips.SBS approach,  at the same privacy cost ($\epsilon=1$) for various data types and learning tasks (image data for grammatical facial expression clustering, mobile-sensing time series data for activity classification, and twitter data to predict topic popularity) with a large number of attributes and cases ($p=77; 117; 301$ and $n=573,820; 216752;27,936$, respectively). Since RON-Gauss  was the proposed method,  it is not surprising that it was the best performer in utility. Compared to the other 3 methods, the performance of  modips varies, depending on the tasks; specifically, it was the second best in classification and clustering, and was the worst in regression.  
\citet{bowen2020comparative} surveyed various dips techniques including the modips.SBS method, compared them conceptually and empirically, and evaluated the statistical utility and inferential properties of the synthetic data via the  techniques through extensive simulation studies. The work employs the results in Theorem \ref{theorem:varcomb} when obtaining inference from multiply synthetic data  ($m=5$). The main conclusions regarding modips.SBS are that with appropriate model specification, it can generate synthetic data offering valid statistical inference (negligible bias and nominal CP) for a practically reasonably small privacy budget and the inferences are often less precise compared to the non-parametric dips that do not consider the sampling variability. 

\citet{liu2020differentially} proposed the DP-ERGM procedure that synthesizes network data via the exponential random graph model (EGRM) in the DP framework. DP-ERGM is a modips procedure that utilizes Algorithm \ref{alg:seapod} to generate posterior samples of ERGM parameters. The work also employed Theorem \ref{theorem:varcomb} when obtaining inference from multiply synthetic networks  ($m=4$). The experiment results suggest that DP-EGRM preserves the original information significantly better than two competitors -- differentially private dyadwise randomized response  and sanitization of the conditional probability of edge given node attribute classes -- in both network statistics and inference of some parametric network models. 

\citet{bowen2021differentially} and \citet{eugenio2021construction} proposed the STEPS and the CIPHER procedures, respectively to generate differentially private  synthetic data to aim for better utility or reduced computational/storage costs. Both methods are model-free in terms of synthesis. Both works obtained the inference in multiply synthetic datasets  ($m=5$) using Theorem \ref{theorem:varcomb} in their experiments.

%---------------------------------------
\vspace{-3pt}\section{Discussion} \label{sec:discussion}\vspace{-9pt}
We propose the modips framework for differentially private data synthesis, along with several procedures to obtain differentially private posterior samples for the implementation of modips. In addition, we propose an inferential combination rule to obtain valid inferences based on multiply synthetic datasets. Our empirical studies demonstrate the validity of the combination rule for inference from differentially private synthetic data, provide insights on the impact of the number of synthetic datasets and privacy budget allocation on inference.

If a dataset is  used mainly for exploratory data analysis or data mining rather than for statistical inference and uncertainty quantification, releasing a single surrogate dataset is workable; otherwise, multiple sets can be released or the sanitization and synthesis mechanisms will need to be modeled when analyzing the synthetic data in order to obtain valid inferences. The former is a more user-friendly approach. Our empirical study suggests that $m\in[3,10]$ is likely to be a proper range for practical use. Existing work that has applied the inferential combination rule  uses $m=4$ or $m=5$. In general, we expect the ``optimal'' $m$, in the sense that the original information preservation is maximized with proper uncertainty quantification at a given privacy budget, varies case by case and  depends on $n$, $p$, the type of sanitizers, among others.  If things are equal, a relatively small $m$ is preferable, as long as it is large enough to capture the between-set variance, so that each synthesis receives a reasonable amount of budget. Small $m$ also helps to save computational/storage costs. We will continue to investigate  theoretically and empirically the choice of $m$ in general settings.

We focus on the modips procedure in the context of the pure $\epsilon$-DP.  Extensions of modips to softer versions of DP that are immune to post processing and closed under composition, such as $(\epsilon,\delta)$-aDP and R\'enyi DP, are straightforward. The only modification in the proposed algorithms for achieving DP that needs is to replace the sanitizer of $\epsilon$-DP with a mechanism that satisfies the softer version of DP.  

We presented modips in the context of the full sample synthesis. It may be possible to extend modips to full population synthesis, which will make an interesting topic for future research, though there will be some technical challenges given the missing values in the unsampled set of a population and the extra sampling step for data release. While it is possible to apply DP in the framework of partial synthesis, we doubt that the robustness and rigor of the privacy guarantees can still be retained in the synthetic data, which arguably one of the biggest advantages of DP over other disclosure risk control approaches. This is because partial synthesis assumes that there is minimal privacy risk from retaining and releasing a subset of the original information (a subset of attributes or individuals), the very idea of which already contradicts the concepts of DP in some sense.

The modips procedure can be challenging for high-dimensional data with a large number of attributes of various types. The difficulty resides in the construction of a parsimonious but representative Bayesian model; identification, and sanitization of sufficient statistics in the case of modips.SBS; and posterior sampling in the high-dimensional setting. Recent development and advancement in efficient Bayesian computation, such as variational Bayes, normalizing flow variational Bayesian, Hamiltonian MC, and sequential MC, can be leveraged in the practical implementation of modips. An alternative is to sanitize the likelihood or the posterior distribution density (or their log versions)  directly if they are bounded while ensuring the sanitized likelihood and posterior distribution density still lead to proper posterior distributions. 

\small
\setlength{\bibsep}{4pt plus 2pt}
\bibliographystyle{apalike}
%\bibliography{myreflist}

\begin{thebibliography}{}

\bibitem[Abadi et~al., 2016]{deep2}
Abadi, M., Chu, A., Goodfellow, I., McMahan, H.~B., Mironov, I., Talwar, K.,
  and Zhang, L. (2016).
\newblock Deep learning with differential privacy.
\newblock {\em arXiv:1607.00133v2}.

\bibitem[Abay et~al., 2018]{abay2018privacy}
Abay, N.~C., Zhou, Y., Kantarcioglu, M., Thuraisingham, B., and Sweeney, L.
  (2018).
\newblock Privacy preserving synthetic data release using deep learning.
\newblock In {\em Joint European Conference on Machine Learning and Knowledge
  Discovery in Databases}, pages 510--526. Springer.

\bibitem[Abowd and Vilhuber, 2008]{abowd2008protective}
Abowd, J.~M. and Vilhuber, L. (2008).
\newblock How protective are synthetic data?
\newblock In {\em Privacy in Statistical Databases}, pages 239--246. Springer.

\bibitem[Acs et~al., 2018]{acs2018differentially}
Acs, G., Melis, L., Castelluccia, C., and De~Cristofaro, E. (2018).
\newblock Differentially private mixture of generative neural networks.
\newblock {\em IEEE Transactions on Knowledge and Data Engineering},
  31(6):1109--1121.

\bibitem[An and Little, 2007]{an2007}
An, D. and Little, R. (2007).
\newblock Multiple imputation: an alternative to top coding for statistical
  disclosure control.
\newblock {\em Journal of the Royal Statistical Society: Series A (Statistics
  in Society)}, 170(4):923--940.

\bibitem[Andr{\'e}s et~al., 2013]{andres2013geo}
Andr{\'e}s, M.~E., Bordenabe, N.~E., Chatzikokolakis, K., and Palamidessi, C.
  (2013).
\newblock Geo-indistinguishability: Differential privacy for location-based
  systems.
\newblock In {\em Proceedings of the 2013 ACM SIGSAC conference on Computer \&
  communications security}, pages 901--914.

\bibitem[Barak et~al., 2007]{barak2007privacy}
Barak, B., Chaudhuri, K., Dwork, C., Kale, S., McSherry, F., and Talwar, K.
  (2007).
\newblock Privacy, accuracy, and consistency too: a holistic solution to
  contingency table release.
\newblock In {\em Proceedings of the twenty-sixth ACM SIGMOD-SIGACT-SIGART
  symposium on Principles of database systems}, pages 273--282. ACM.

\bibitem[Bernardo and Smith, 1994]{Bernardo1994}
Bernardo, J. and Smith, A. (1994).
\newblock {\em Bayesian Theory}.
\newblock Wiley.

\bibitem[Bindschadler, 2018]{bindschadler2018privacy}
Bindschadler, V. (2018).
\newblock {\em Privacy-preserving seedbased data synthesis}.
\newblock PhD thesis, University of Illinois at Urbana-Champaign.

\bibitem[Bindschaedler et~al., 2017]{bindschaedler2017plausible}
Bindschaedler, V., Shokri, R., and Gunter, C.~A. (2017).
\newblock Plausible deniability for privacy-preserving data synthesis.
\newblock {\em arXiv preprint arXiv:1708.07975}.

\bibitem[Blackwell and Ramamoorthi, 1982]{Blackwell1982}
Blackwell, D. and Ramamoorthi, R.~V. (1982).
\newblock A bayes but not classically sufficient statistic.
\newblock {\em Annals of Statistics}, 10(3):1025--1026.

\bibitem[Blei et~al., 2017]{blei2017variational}
Blei, D.~M., Kucukelbir, A., and McAuliffe, J.~D. (2017).
\newblock Variational inference: A review for statisticians.
\newblock {\em Journal of the American statistical Association},
  112(518):859--877.

\bibitem[Blum et~al., 2008]{blum2008learning}
Blum, A., Ligett, K., and Roth, A. (2008).
\newblock A learning theory approach to non-interactive database privacy.
\newblock In {\em Proceedings of the fortieth annual ACM symposium on Theory of
  computing}, pages 609--618. ACM.

\bibitem[Bowen, 2018]{bowen}
Bowen, C.~M. (2018).
\newblock {\em Data Privacy via Integration of Differential Privacy and Data
  Synthesis}.
\newblock PhD thesis, University of Notre Dame.

\bibitem[Bowen and Liu, 2020]{bowen2020comparative}
Bowen, C.~M. and Liu, F. (2020).
\newblock Comparative study of differentially private data synthesis methods.
\newblock {\em Statistical Science}, 35(2):280--307.

\bibitem[Bowen et~al., 2021]{bowen2021differentially}
Bowen, C.~M., Liu, F., and Su, B. (2021).
\newblock Differentially private data release via statistical election to
  partition sequentially.
\newblock {\em METRON}, pages 1--31.

\bibitem[Bun et~al., 2018]{bun2018composable}
Bun, M., Dwork, C., Rothblum, G.~N., and Steinke, T. (2018).
\newblock Composable and versatile privacy via truncated cdp.
\newblock In {\em Proceedings of the 50th Annual ACM SIGACT Symposium on Theory
  of Computing}, pages 74--86.

\bibitem[Burgette and Reiter, 2013]{Burgette2013}
Burgette, L.~F. and Reiter, J.~P. (2013).
\newblock Multiple-shrinkage multinomial probit models with applications to
  simulating geographies in public use data.
\newblock {\em Bayesian Analysis}, 8(2):453--478.

\bibitem[Caiola and Reiter, 2010]{ReiterRandomForest2010}
Caiola, G. and Reiter, J.~P. (2010).
\newblock Random forests for generating partially synthetic, categorical data.
\newblock {\em Transactions on Data Privacy}, 3(1):27 -- 42.

\bibitem[Chanyaswad et~al., 2019]{RON-Gauss2019}
Chanyaswad, T., Liu, C., and Mittal, P. (2019).
\newblock Ron-gauss: Enhancing utility in non-interactive private data release.
\newblock In {\em Proceedings on Privacy Enhancing Technologies (PETS)}.

\bibitem[Charest, 2010]{Charest2010}
Charest, A.~S. (2010).
\newblock How can we analyze differentially private synthetic datasets.
\newblock {\em Journal of Privacy and Confidentiality}, 2(2):Article 3.

\bibitem[Chaudhuri et~al., 2011]{Chaudhuri2011}
Chaudhuri, K., Monteleoni, C., and Sarwate, A.~D. (2011).
\newblock Differentially private empirical risk minimization.
\newblock {\em JMLR: Workshop and Conference Proceedings}, 12:1069--1109.

\bibitem[Chaudhuri et~al., 2013]{Chaudhuri2012PCA}
Chaudhuri, K., Sarwate, A., and Sinha, K. (2013).
\newblock A near-optimal differentially private principal components.
\newblock {\em The Journal of Machine Learning Research}, 14:2905--2943.

\bibitem[Dimitrakakis et~al., 2014]{dimitrakakis2014robust}
Dimitrakakis, C., Nelson, B., Mitrokotsa, A., and Rubinstein, B.~I. (2014).
\newblock Robust and private bayesian inference.
\newblock In {\em International Conference on Algorithmic Learning Theory},
  pages 291--305. Springer.

\bibitem[Domingo-Ferrer and Sayg\`{z}n, 2008]{Domingobook2008}
Domingo-Ferrer, J. and Sayg\`{z}n, Y., editors (2008).
\newblock {\em Privacy in statistical database}.
\newblock Springer-Verlag Berlin Heidelberg.

\bibitem[Domingo-Ferrer and Torra, 2001]{Domingo2001}
Domingo-Ferrer, J. and Torra, V. (2001).
\newblock Disclosure control methods and information loss for microdata.
\newblock In Doyle, P., Lane, J., Theeuwes, J., and Zayatz, L., editors, {\em
  Confidentiality, disclosure, and data access : Theory and practical
  applications for statistical agencies}, pages 91--110. Elsevier.

\bibitem[Domingo-Ferrer and Torra, 2004]{Domingo2004}
Domingo-Ferrer, J. and Torra, V. (2004).
\newblock Disclosure risk assessment in statistical data protection.
\newblock {\em Journal of Computational and Appled Mathematics},
  164-165(1):285--293.

\bibitem[Dong et~al., 2019]{dong2019gaussian}
Dong, J., Roth, A., and Su, W.~J. (2019).
\newblock Gaussian differential privacy.
\newblock {\em arXiv preprint arXiv:1905.02383}.

\bibitem[Drechsler and Reiter, 2011]{rdrechsler2011paper}
Drechsler, J. and Reiter, J.~P. (2011).
\newblock An empirical evaluation of easily implemented, nonparametric methods
  for generating synthetic data sets.
\newblock {\em Computational Statistics and Data Analysis}, 55(12):461--468.

\bibitem[Dwork, 2006]{dwork2006}
Dwork, C. (2006).
\newblock Differential privacy.
\newblock In {\em Proceedings of the International Colloqium on Automata,
  Languages and Programming (ICALP)}, pages 1--12. Springer-Verlag ARCoSS.

\bibitem[Dwork et~al., 2006a]{dwork2006delta}
Dwork, C., Kenthapadi, K., McSherry, F., Mironov, I., and Naor, M. (2006a).
\newblock Our data, ourselves: privacy via distributed noise generation.
\newblock In {\em Advances in Cyptology: Proceedings of EUROCRYPT}, pages
  485--503. Springer Berlin Heidelberg.

\bibitem[Dwork et~al., 2006b]{dwork2006calibrating}
Dwork, C., McSherry, F., Nissim, K., and Smith, A. (2006b).
\newblock Calibrating noise to sensitivity in private data analysis.
\newblock In {\em Theory of cryptography}, pages 265--284. Springer.

\bibitem[Dwork and Roth, 2014a]{privacybook}
Dwork, C. and Roth, A. (2014a).
\newblock {\em The Algorithmic Foundation of Differential Privacy}.
\newblock Now Publishes, Inc.

\bibitem[Dwork and Roth, 2014b]{dwork2014algorithmic}
Dwork, C. and Roth, A. (2014b).
\newblock The algorithmic foundations of differential privacy.
\newblock {\em Foundations and Trends in Theoretical Computer Science},
  9(3-4):211--407.

\bibitem[Dwork and Rothblum, 2016]{cPD}
Dwork, C. and Rothblum, G.~N. (2016).
\newblock Concentrated differential privacy.
\newblock {\em arXiv:1603.01887v2}.

\bibitem[Dwork et~al., 2010]{dwork2010boosting}
Dwork, C., Rothblum, G.~N., and Vadhan, S. (2010).
\newblock Boosting and differential privacy.
\newblock In {\em 2010 IEEE 51st Annual Symposium on Foundations of Computer
  Science}, pages 51--60. IEEE.

\bibitem[Eugenio, 2019]{eugenio}
Eugenio, E.~C. (2019).
\newblock {\em Some Methods for Differentially Private Data Synthesis}.
\newblock PhD thesis, University of Notre Dame.

\bibitem[Eugenio and Liu, 2021]{eugenio2021construction}
Eugenio, E.~C. and Liu, F. (2021).
\newblock Construction of microdata from a set of differentially private
  low-dimensional contingency tables through solving linear equations with
  tikhonov regularization.
\newblock {\em Advances in Intelligent Systems and Computing: Proceedings of
  Computing Conference 2021, London, IK}.

\bibitem[Fienberg et~al., 1997]{fienberg1997bayesian}
Fienberg, S.~E., Makov, U.~E., and Sanil, A.~P. (1997).
\newblock A bayesian approach to data disclosure: Optimal intruder behavior for
  continuous data.
\newblock {\em Journal of Official Statistics}, 13:75--79.

\bibitem[Goodfellow et~al., 2014]{goodfellow2014generative}
Goodfellow, I.~J., Pouget-Abadie, J., Mirza, M., Xu, B., Warde-Farley, D.,
  Ozair, S., Courville, A., and Bengio, Y. (2014).
\newblock Generative adversarial networks.
\newblock {\em arXiv preprint arXiv:1406.2661}.

\bibitem[Hall et~al., 2013a]{function}
Hall, R., Rinaldo, A., and Wasserma, L. (2013a).
\newblock Differential privacy for functions and functional data.
\newblock {\em Journal of Machine Learning Research}, 14:703--727.

\bibitem[Hall et~al., 2013b]{hall2013differential}
Hall, R., Rinaldo, A., and Wasserman, L. (2013b).
\newblock Differential privacy for functions and functional data.
\newblock {\em The Journal of Machine Learning Research}, 14(1):703--727.

\bibitem[Hall et~al., 2012]{randomDP}
Hall, R., Rinaldoy, A., and Wasserman, L. (2012).
\newblock Random differential privacy.
\newblock {\em Journal of Privacy and Confidentiality}, 4(2):43--59.

\bibitem[Hardt et~al., 2012]{multiplicative}
Hardt, M., Ligett, K., and McSherry, F. (2012).
\newblock A simple and practical algorithm for differentially private data
  release.
\newblock {\em arXiv:1012.4763v2}.

\bibitem[He et~al., 2015]{he2015}
He, X., Cormode, G., Machanavajjhala, A., Procopiuc, C.~M., and Srivastava, D.
  (2015).
\newblock Dpt: Differentially private trajectory synthesis using hierarchical
  reference systems.
\newblock {\em Proceedings of the VLDB Endowment}, 8(11):1154--1165.

\bibitem[J{\"a}lk{\"o} et~al., 2016]{jalko2016differentially}
J{\"a}lk{\"o}, J., Dikmen, O., and Honkela, A. (2016).
\newblock Differentially private variational inference for non-conjugate
  models.
\newblock {\em arXiv preprint arXiv:1610.08749}.

\bibitem[Jordon et~al., 2018]{jordon2018pate}
Jordon, J., Yoon, J., and Van Der~Schaar, M. (2018).
\newblock Pate-gan: Generating synthetic data with differential privacy
  guarantees.
\newblock In {\em International Conference on Learning Representations}.

\bibitem[Kang et~al., 2020]{kang2020study}
Kang, J., Jeong, S., Hong, D., and Seo, C. (2020).
\newblock A study on synthetic data generation based safe differentially
  private gan.
\newblock {\em Journal of the Korea Institute of Information Security \&
  Cryptology}, 30(5):945--956.

\bibitem[Karwa et~al., 2016]{jrssc}
Karwa, V., Krivitsky, P.~N., and Slavkovi\'{c}, A.~B. (2016).
\newblock Sharing social network data: differentially private estimation of
  exponential family random-graph models.
\newblock {\em Applied Statistics (JRSS-C)}, page DOI: 10.1111/rssc.12185.

\bibitem[Karwa and Slavkovi\'{c}, 2015]{annals}
Karwa, V. and Slavkovi\'{c}, A.~B. (2015).
\newblock Inference using noisy degrees: differentially private $\beta$-model
  and synthetic graphs.
\newblock {\em Annals of Statistics}, 44 (1):87--112.

\bibitem[Kifer and Lin, 2012]{kifer2012axiomatic}
Kifer, D. and Lin, B.-R. (2012).
\newblock An axiomatic view of statistical privacy and utility.
\newblock {\em Journal of Privacy and Confidentiality}, 4(1).

\bibitem[Kifer and Machanavajjhala, 2011]{kifer2011no}
Kifer, D. and Machanavajjhala, A. (2011).
\newblock No free lunch in data privacy.
\newblock In {\em Proceedings of the 2011 ACM SIGMOD International Conference
  on Management of data}, pages 193--204.

\bibitem[Kifer et~al., 2012]{Kifer2012}
Kifer, D., Smith, A., and Thakurta, A. (2012).
\newblock Private convex empirical risk minimization and high-dimensional
  regression.
\newblock {\em JMLR: Workshop and Conference Proceedings}, 23:25.1--25.40.

\bibitem[Lei, 2011]{m-estimator}
Lei, J. (2011).
\newblock Differentially private m-estimators.
\newblock {\em Proceedings of Advances in Neural Information Processing
  Systems}.

\bibitem[Li et~al., 2016]{li2016differentially}
Li, D., Zhang, W., and Chen, Y. (2016).
\newblock Differentially private network data release via stochastic kronecker
  graph.
\newblock In {\em International Conference on Web Information Systems
  Engineering}, pages 290--297. Springer.

\bibitem[Li et~al., 2014]{copula}
Li, H., Xiong, L., and Jiang, X. (2014).
\newblock Differentially private synthesization of multi-dimensional data using
  copula functions.
\newblock {\em Advances in Database Technology}, pages 475--486.

\bibitem[Little, 1993]{little1993}
Little, R. (1993).
\newblock Statistical analysis of masked data.
\newblock {\em Journal of the Official Statistics}, 9:407--407.

\bibitem[Little et~al., 2004]{Liu2004}
Little, R., Liu, F., and Raghunathan, T. (2004).
\newblock Statistical disclosure techniques based on multiple imputation.
\newblock In Gelman, A. and Meng, X.-L., editors, {\em Applied Bayesian
  Modeling and Causal Inference from Incomplete-Data Perspectives: An essential
  journey with Donald Rubin's statistical family}, page Chapter II.13. John
  Wiley \& Sons.

\bibitem[Liu, 2016]{liu2016model}
Liu, F. (2016).
\newblock Model-based differentially private data synthesis.
\newblock {\em arXiv preprint arXiv:1606.08052}.

\bibitem[Liu, 2019a]{ggm}
Liu, F. (2019a).
\newblock Generalized gaussian mechanism for differential privacy.
\newblock {\em IEEE Transactions on Knowledge and Data Engineering},
  31(4):747--756.

\bibitem[Liu, 2019b]{Liu2019}
Liu, F. (2019b).
\newblock Statistical properties of sanitized results from differentially
  private laplace mechanism with univariate bounding constraints.
\newblock {\em Transactions on Data Privacy}, 12:169--195.

\bibitem[Liu et~al., 2020]{liu2020differentially}
Liu, F., Eugenio, E., Jin, I.~H., and Bowen, C. (2020).
\newblock Differentially private generation of social networks via exponential
  random graph models.
\newblock In {\em 2020 IEEE 44th Annual Computers, Software, and Applications
  Conference (COMPSAC)}, pages 1695--1700. IEEE.

\bibitem[Liu and Little, 2002]{Liu2002}
Liu, F. and Little, R. (2002).
\newblock Selective multiple imputation of keys for statistical disclosure
  limitation in microdata.
\newblock {\em Proceedings of 2002 American Statistical Association Joint
  Statistical Meeting}.

\bibitem[Machanavajjhala et~al., 2008]{onthemap}
Machanavajjhala, A., Kifer, D., Abowd, J., Gehrke, J., and Vilhuber, L. (2008).
\newblock Privacy: Theory meets practice on the map.
\newblock {\em IEEE ICDE IEEE 24th International Conference}, pages 277 -- 286.

\bibitem[Manrique-Vallier and Reiter, 2012]{manrique2012estimating}
Manrique-Vallier, D. and Reiter, J.~P. (2012).
\newblock Estimating identification disclosure risk using mixed membership
  models.
\newblock {\em Journal of the American Statistical Association},
  107(500):1385--1394.

\bibitem[McClure and Reiter, 2012]{mcclure2012differential}
McClure, D. and Reiter, J.~P. (2012).
\newblock Differential privacy and statistical disclosure risk measures: An
  investigation with binary synthetic data.
\newblock {\em Transactions on Data Privacy}, 5(3):535--552.

\bibitem[McSherry and Talwar, 2007]{mcsherry2007mechanism}
McSherry, F. and Talwar, K. (2007).
\newblock Mechanism design via differential privacy.
\newblock In {\em Foundations of Computer Science, 48-th Annual IEEE Symposium,
  FOCS'07}, pages 94--103. IEEE.

\bibitem[Melis, 2018]{melis2018building}
Melis, L. (2018).
\newblock {\em Building and evaluating privacy-preserving data processing
  systems}.
\newblock PhD thesis, UCL (University College London).

\bibitem[Meng, 1994]{meng1994multiple}
Meng, X.-L. (1994).
\newblock Multiple-imputation inferences with uncongenial sources of input.
\newblock {\em Statistical Science}, pages 538--558.

\bibitem[Mironov, 2017]{mironov2017renyi}
Mironov, I. (2017).
\newblock R{\'e}nyi differential privacy.
\newblock In {\em 2017 IEEE 30th Computer Security Foundations Symposium
  (CSF)}, pages 263--275. IEEE.

\bibitem[Nogales et~al., 2000]{Nogales2000}
Nogales, A., Oyola, J., and Perez, P. (2000).
\newblock On conditional independence and the relationship between sufficiency
  and invariance under the bayesian point of view.
\newblock {\em Statistics \& Probability Letters}, 46(1):75--84.

\bibitem[Proserpio et~al., 2012]{graph}
Proserpio, D., Goldberg, S., and McSherry, F. (2012).
\newblock A workflow for differentially-private graph synthesis.
\newblock {\em Proceedings of the 2012 ACM workshop on online social networks},
  pages 13--18.

\bibitem[Quick, 2019]{quick2019generating}
Quick, H. (2019).
\newblock Generating poisson-distributed differentially private synthetic data.
\newblock {\em arXiv preprint arXiv:1906.00455}.

\bibitem[Raab et~al., 2017]{raab2017practical}
Raab, G.~M., Nowok, B., and Dibben, C. (2017).
\newblock Practical data synthesis for large samples.
\newblock {\em Journal of Privacy and Confidentiality}, 7:4.

\bibitem[Raghunathan et~al., 2003]{raghunathan2003multiple}
Raghunathan, T.~E., Reiter, J.~P., and Rubin, D.~B. (2003).
\newblock Multiple imputation for statistical disclosure limitation.
\newblock {\em Journal of official Statistics}, 19(1):1--16.

\bibitem[Reiter, 2005a]{ReiterCART2005}
Reiter, J. (2005a).
\newblock Using cart to generate partially synthetic public use microdata.
\newblock {\em Journal of Official Statistics}, 21:441--462.

\bibitem[Reiter, 2003]{reiter2003}
Reiter, J.~P. (2003).
\newblock Inference for partially synthetic, public use microdata sets.
\newblock {\em Survey Methodology}, 29(2):181--188.

\bibitem[Reiter, 2005b]{reiter2005estimating}
Reiter, J.~P. (2005b).
\newblock Estimating risks of identification disclosure in microdata.
\newblock {\em Journal of the American Statistical Association},
  100(472):1103--1112.

\bibitem[Reiter et~al., 2014]{reiter2014bayesian}
Reiter, J.~P., Wang, Q., and Zhang, B. (2014).
\newblock Bayesian estimation of disclosure risks for multiply imputed,
  synthetic data.
\newblock {\em Journal of Privacy and Confidentiality}, 6(1):2.

\bibitem[Rezende and Mohamed, 2015]{rezende2015variational}
Rezende, D. and Mohamed, S. (2015).
\newblock Variational inference with normalizing flows.
\newblock In {\em International Conference on Machine Learning}, pages
  1530--1538. PMLR.

\bibitem[Rubin, 1987]{MIbook}
Rubin, D.~B. (1987).
\newblock {\em Multiple Imputation for Nonresponse in Surveys}.
\newblock John Wiley \& Sons, New York.

\bibitem[Rubin, 1993]{rubin1993statistical}
Rubin, D.~B. (1993).
\newblock Statistical disclosure limitation.
\newblock {\em Journal of official Statistics}, 9(2):461--468.

\bibitem[Shokri and Shmatikov, 2015]{deep1}
Shokri, R. and Shmatikov, V. (2015).
\newblock Privacy-preserving deep learning.
\newblock {\em ACM CCS}, pages 1310--1321.

\bibitem[Smith, 2011]{estimation}
Smith, A. (2011).
\newblock Privacy-preserving statistical estimation with optimal convergence
  rates.
\newblock {\em Proceeding of STOC '11 Proceedings of the forty-third annual ACM
  symposium on Theory of computing}, pages 813--822.

\bibitem[Smith and Thakurta, 2013]{lasso}
Smith, A. and Thakurta, A. (2013).
\newblock Differentially private model selection via stability arguments and
  the robustness of the lasso.
\newblock {\em JMLR: Workshop and Conference Proceedings}, 30:1–32.

\bibitem[Waites and Cummings, 2020]{waites2020differentially}
Waites, C. and Cummings, R. (2020).
\newblock Differentially private normalizing flows for privacy-preserving
  density estimation.
\newblock In {\em ICML Workshop on Invertible Neural Networks, Normalizing
  Flows, and Explicit Likelihood Models}.

\bibitem[Wang and Wu, 2013]{wang2013preserving}
Wang, Y. and Wu, X. (2013).
\newblock Preserving differential privacy in degree-correlation based graph
  generation.
\newblock {\em Transactions on Data Privacy}, 6:127–145.

\bibitem[Wang et~al., 2015]{wang2015privacy}
Wang, Y.-X., Fienberg, S., and Smola, A. (2015).
\newblock Privacy for free: Posterior sampling and stochastic gradient monte
  carlo.
\newblock In {\em International Conference on Machine Learning}, pages
  2493--2502.

\bibitem[Wasserman and Zhou, 2010]{wasserman2010statistical}
Wasserman, L. and Zhou, S. (2010).
\newblock A statistical framework for differential privacy.
\newblock {\em Journal of the American Statistical Association},
  105(489):375--389.

\bibitem[Xiao et~al., 2014]{xiao2014differentially}
Xiao, Q., Chen, R., and Tan, K.-L. (2014).
\newblock Differentially private network data release via structural inference.
\newblock In {\em Proceedings of the 20th ACM SIGKDD international conference
  on Knowledge discovery and data mining}, pages 911--920.

\bibitem[Yu et~al., 2014]{Yu2014}
Yu, F., Fienberg, S.~E., Slavkovic, A.~B., and Uhler, C. (2014).
\newblock Scalable privacy-preserving data sharing methodology for genome-wide
  association studies.
\newblock {\em Journal of Biomedical Informatics}, 50:133--141.

\bibitem[Zhang et~al., 2014]{privbayes}
Zhang, J., Cormode, G., Procopiuc, C.~M., Srivastava, D., and Xiao, X. (2014).
\newblock Privbayes: Private data release via bayesian networks.
\newblock In {\em Proceedings of the 2014 ACM SIGMOD international conference
  on Management of data}, pages 1423--1434. ACM.

\end{thebibliography}

\vspace{-9pt}
\section*{Appendix}\vspace{-9pt}
\appendix
\numberwithin{equation}{section}
\section{\large{Proof of Proposition \ref{pro:dp}}}\vspace{-9pt}
\begin{proof}
\noindent$\Pr(\tilde{\x}^{*(i)} \!\in \!Q|\x)\!=\! \E(\E(\mathrm{I}(\tilde{\x}^{*(i)} \!\in \! Q)|\bs{\theta}^{*(i)})|\x)\!=\!
\int \!\!\int \mathrm{I}(\tilde{\x}^{*(i)} \!\in \!Q) f(\tilde{\x}^{*(i)}|\bs{\theta}^{*(i)} )f(\bs{\theta}^{*(i)}|\x)d\tilde{\x}^{*(i)}d\bs{\theta}^{*(i)}\\
\!=\!\int\!\!\int \mathrm{I}(\tilde{\x}^{*(i)} \!\in \! Q) f(\tilde{\x}^{*(i)}|\bs{\theta}^{*(i)}) \frac{f(\bs{\theta}^{*(i)}|\x)}{f(\bs{\theta}^{*(i)}|\x')}f(\bs{\theta}^{*(i)}|\x')d\tilde{\x}^{*(i)}d\bs{\theta}^{*(i)}$.
Since $\bs{\theta}^{*(i)}$ is of $\epsilon'\!=\!(\epsilon-\epsilon_0)/m$-DP, 
that is, $e^{-\epsilon'}\!\!\le\!\frac{f(\bs{\theta}^{*(i)}|\x)}{f(\bs{\theta}^{*(i)}|\x')}\!\le\! e^{\epsilon'}$,  $e^{-\epsilon'}\!\int\!\! \int\mathrm{I}(\tilde{\x}^{*(i)}\in Q) f(\tilde{\x}^{*(i)}|\bs{\theta}^{*(i)})f(\bs{\theta}^{*(i)}|\x')d\tilde{\x}^{*(i)}d\bs{\theta}^{*(i)}\!\le\!\int \!\!\int \mathrm{I}(\tilde{\x}^{*(i)}\!\in\!Q)\\ \frac{f(\bs{\theta}^{*(i)}|\x)}{f(\bs{\theta}s^{*(i)}|\x')}  f(\bs{\tilde{\x}^{*(i)}|\bs{\theta}^{*(i)}})f(\bs{\theta}^{*(i)}|\x')d\tilde{\x}^{*(i)}d\bs{\theta}^{*(i)}
\!\le\! e^{\epsilon'}\!\int\!\!\int\mathrm{I}(\tilde{\x}^{*(i)}\in Q) f(\tilde{\x}^{*(i)}|\bs{\theta}^{*(i)})f(\bs{\theta}^{*(i)}|\x')d\tilde{\x}^{*(i)}d\bs{\theta}^{*(i)}$. \\ 
Given that
$\int \!\!\int\mathrm{I}(\tilde{\x}^{*(i)}\!\in\! Q) f(\tilde{\x}^{*(i)}|\bs{\theta}^{*(i)})f(\bs{\theta}^{*(i)}|\x')d\tilde{\x}^{*(i)}d\bs{\theta}^{*(i)}\!=\!\E(\E(\mathrm{I}(\tilde{\x}^{*(i)}\in Q)|\s^{*(i)})|\x') \!=\!\Pr(\tilde{\x}^{*(i)}\!\in\!
Q|\x')$, then $e^{-\epsilon'}\Pr(\tilde{\x}^{*(i)}\in Q|\x') \le \Pr(\tilde{\x}^{*(i)}\in Q|\x)\le e^{\epsilon}\Pr(\tilde{\x}^{*(i)}\in Q|\x')$, and $\tilde{\x}^{*(i)}$ is released with $\epsilon'$-DP. Following the sequential composition principle, synthesizing $m$ sets of $m\epsilon'=\epsilon-\epsilon_0$ DP; plus the budget spent on model selection $\epsilon_0$, releasing $m$ sets of synthetic data is of $\epsilon$-DP.\vspace{-6pt}
\end{proof}

%------------------
\section{\large{Proof of Proposition \ref{pro.SBS}}}\vspace{-9pt}
\begin{proof} In the $i$-iteration for $i=1,\ldots,m$,
$\Pr(\bs{\theta}^{*(i)}\in Q|\x) = \E(\E(\mathrm{I}(\bs{\theta}^{*(i)}\!\in\! Q)|\s^{*(i)})|\x)\!=\!
\int \!\!\int \mathrm{I}(\bs{\theta}^{*(i)}\!\in\!Q)f(\bs{\theta}^{*(i)}|\s^{*(i)})f(\s^{*(i)}|\x)d\bs{\theta}^{*(i)}d\s^{*(i)}\!=\!\!
\int\!\!\int \mathrm{I}(\bs{\theta}^*\!\in\! Q) f(\bs{\theta^*|\s^*}) \frac{f(\s^*|\x)}{f(\s^{*(i)}|\x')}f(\s^{*(i)}|\x')d\bs{\theta}^{*(i)}d\s^{*(i)}$. $\s^{*(i)}$ is of $\epsilon'$-DP; therefore, $e^{-\epsilon'}\!\le\!\frac{f(\s^{*(i)}|\x)}{f(\s^{*(i)}|\x')}\!\le\! e^{\epsilon'}$ and $e^{-\epsilon'}\!\int\!\!\int\mathrm{I}(\bs{\theta}^{*(i)}\!\in\! Q) f(\bs{\theta^{*(i)}|\s^{*(i)}})
f(\s^{*(i)}|\x')d\bs{\theta}^{*(i)}d\s^{*(i)}\!\le\!
\int \!\!\int \mathrm{I}(\bs{\theta}^{*(i)}\!\in\! Q)  \frac{f(\s^{*(i)}|\x)}{f(\s^{*(i)}|\x')} f(\bs{\theta^{*(i)}|\s^{*(i)}})f(\s^{*(i)}|\x)d\bs{\theta}^{*(i)}d\s^{*(i)}
\!\le\! e^{\epsilon'}\!\int\!\!\int\mathrm{I}(\bs{\theta}^{*(i)} \!\in\! Q) f(\bs{\theta^{*(i)}|\s^{*(i)}})f(\s^{*(i)}|\x') \\
d\bs{\theta}^{*(i)}d\s^{*(i)}$. 
Since $\int\!\!\int\mathrm{I}(\bs{\theta}^{*(i)}\!\in\! Q) f(\bs{\theta^{*(i)}|\s^{*(i)}})f(\s^{*(i)}|\x)
d\bs{\theta}^{*(i)}d\s^{*(i)}=\E(\E(\mathrm{I}(\bs{\theta}^{*(i)}\!\in\! Q)|\s^{*(i)})|\x) =\Pr(\bs{\theta}^{*(i)}\!\in\! Q|\x)$, similarly when $\x$ is replaced by $\x'$, then  $e^{-\epsilon}\Pr(\bs{\theta}^{*(i)}
\in Q|\x') \le \Pr(\bs{\theta}^{*(i)}\in Q|\x)\le  e^{\epsilon'}\Pr(\bs{\theta}^{*(i)}\in Q|\x')$,  $e^{-\epsilon'}\le \frac{\Pr(\bs{\theta}^{*(i)}\in Q|\x)}{\Pr(\bs{\theta}^{*(i)}\in Q|\x')}\le  e^{\epsilon'}$, and a random sample $\bs{\theta}^{*(i)}$ also satisfies $\epsilon'$-DP. The rest of the proof is the same as the proof for Theorem \ref{pro:dp}, leading to the final conclusion that the sanitized data $\x$ from the modips.SBS satisfies DP.\vspace{-6pt}
\end{proof}

%------------------
\section{\large Proof of Proposition \ref{pro:Deltau}}\vspace{-9pt}
\begin{proof}
$u_i\!=\!-\log\big(\int_{\bs\theta\in \mathcal{B}_i}\! f(\bs\theta|\x)d\bs\theta\big)\!=\! -(\log(\bar{f}_i(\bs\theta|\x))+\log(V_i))$ and $|u_i(\x)-u(\x')|\!=\! |\log(\bar{f}_i(\bs\theta|\x))-\log(\bar{f}_i(\bs\theta|\x'))|<2A$. Therefore, $\Delta_u=\max_{\mathcal{B},\bs\theta,d(\x,\x')=1}|u_i(\x)-u(\x')|<2A$.\vspace{-6pt}
\end{proof}

\section{\large Proof of Proposition \ref{pro:GS1}}\vspace{-9pt}
\begin{proof}
Draw $N$ samples from $f(\bs\theta|\x)$ and $f(\bs\theta|\x')$, where  $\x$ and $\x'$ are a pair of neighboring datasets, and form the histograms based on the $N$ samples, respectively.  The two histograms share the same bin cut points. Denote the number of bins in the $j$-th marginal histogram by $B_j$ for $j=1\ldots,p$, and the width of the bins  by $\mathbf{h}_j=(h_{1,j},\ldots,h_{B_j,j})$. The total number of bins in each histogram is  $B\!=\!\prod_{j=1}^p\!B_j$, and the bin count of bin $\mathcal{B}_i$ in the two histograms by $N_i$ and $N'_i$, respectively.  The maximum change in the log(proportion) of bin $\mathcal{B}_i$  given one-individual change from $\x$ to $\x'$ is $\Delta_1\!=\!\max\{|\log(N_i/N)\!-\!\log(N'_i/N)|\}$. Since $N_i/N\!\approx\!\bar{f}_{\bs\theta\in\mathbf{B_i}}(\bs\theta|\x) \prod_{j=1}^{p}\!h_{j(i),j}$ and $N'_i/N\!\approx\! \bar{f}_{\bs\theta\in\mathbf{B_i}}(\bs\theta|\x') \prod_{j=1}^p\!h_{j(i),j}$, $\Delta_1= \max\{|\log(\bar{f}_{\bs\theta\in\mathbf{B_i}}(\bs\theta|\x)\!-\!\log(\bar{f}_{\bs\theta\in\mathbf{B_i}}(\bs\theta|\x'))|\}\!<\!2A$.
\vspace{-6pt} \end{proof}

%----------------------
\section{\large Proof of Proposition \ref{pro:GS2}}\vspace{-9pt}
\begin{proof}
The proof is similar to that for Proposition \ref{pro:GS1} except for the last step.  Let $\prod_{j=1}^{p}h_{j(i),j}=V_i$. Since $N_i\approx N\bar{f}_{\bs\theta\in\mathbf{B_i}}(\bs\theta|\x)V_i$ and $N'_i\approx N\bar{f}_{\bs\theta\in\mathbf{B_i}}(\bs\theta|\x')V_i$,  $\Delta_i=
\max\{|N_i-N'_i|\}\approx NV_i\\ \max\{\bar{f}_{\bs\theta\in\mathbf{B_i}}(\x)-\bar{f}_{\bs\theta\in\mathbf{B_i}}(\x')\}<NV_i\max\{\bar{f}_{\bs\theta\in\mathbf{B_i}}(\x),\bar{f}_{\bs\theta\in\mathbf{B_i}}(\x')\}<NV_iG$.\vspace{-6pt}
\end{proof}

%------------------
\section{\large{Proof of  Theorem \ref{theorem:varcomb}}}\label{app:varcomb}\vspace{-9pt}

\textbf{Part a).}  The likelihood is $f(\x|\bs\theta)$. Synthetic data $\tilde{\x}^*$ via the modips procedure is generated from $f(\X|\bs\theta^{*})$, where $\bs\theta^{*(i)}$ is a random sample from the sanitized posterior distribution  $f^*(\bs\theta|\x)$. We assume that  $f^*(\bs\theta|\x)$ is consistent for $f(\bs\theta|\x)$.   WLOS, suppose $\theta$ is a scalar. If the estimator of $\hat{\theta}$ based on original $\x$ is consistent (e.g., MLE, posterior mean) for the parameter of interest $\theta$, then the same estimator $\hat{\theta}^*$ but based on the sanitized $\tilde\x^*$ is consistent for $\theta^*$ as the distribution that generates $\x$ and  $\tilde{\x}^*$ are the same except the underlying parameter values.  The mean squared error of $\hat{\theta}^*$ as an estimate for $\theta$ is
\begin{align}\vspace{-6pt}
&\E_{\X^*}(\hat{\theta}^*-\theta)^2
=\E_{\X^*}(\hat{\theta}^*-\theta^*+\theta^*-\theta)^2\notag\\
=\;&\E_{\X^*}(\hat{\theta}^*-\theta^*)^2+\E_{\X^*}(\theta^*-\theta)^2+
2\E_{\X^*}[(\hat{\theta}^*-\theta^*)(\theta^*-\theta)]\notag\\
=\;&\E_{\X^*}(\hat{\theta}^*-\theta^*)^2+(\theta^*-\theta)^2+
2(\theta^*-\theta)\E_{\X^*}(\hat{\theta}^*-\theta^*)\label{eqn:3terms}\\
\rightarrow&\; (\theta^*-\theta)^2 \mbox{ as } n\rightarrow\infty, \mbox{ which }
\rightarrow\;0 \mbox{ as } \epsilon\rightarrow\infty.\notag
\vspace{-6pt}\end{align}
The first and third terms $\E_{\X^*}(\hat{\theta}^*-\theta^*)^2$ and $\E_{\X^*}(\hat{\theta}^*-\theta^*)\rightarrow0$  as $n\rightarrow\infty$ in Eqn (\ref{eqn:3terms}) due to consistency of $\hat{\theta}^*$ for $\theta^*$.  As for the second term $(\theta^*-\theta)^2$, $\theta^*$ is a draw from the sanitized posterior distribution $f^*(\theta|\x)$. If there is no sanitization, $f(\theta|\x)$ approaches a degenerate distribution at point $\theta$  as $n\rightarrow\infty$; in other words, $f(\theta|\x)\overset{d}{\rightarrow}\theta$ as $n\rightarrow\infty$.  With sanitization, since $f^*(\theta|\x)\overset{d}{\rightarrow}f(\theta|\x)$ as $\epsilon\rightarrow\infty$ per consistency (definition \ref{def:consistency}) and  $f(\theta|\x)\overset{d}{\rightarrow}\theta$ as $n\!\rightarrow\!\infty$, then   $f^*(\theta|\x)\overset{d}{\rightarrow}\theta$ as $n\!\rightarrow\!\infty$ and $\epsilon\!\rightarrow\!\infty$. Taken together, $(\theta^*\!-\!\theta)^2 \rightarrow0$ as $n\!\rightarrow\!\infty$ and $\epsilon\!\rightarrow\!\infty$. 

It is possible the parameter of inference interest is not $\theta$ per se, the parameter involved in the sanitization and synthesis on the data curator's side, but a different parameter, say $\beta$. Assuming the distribution or model assumed by the data analyst for the released sanitized data is congenial in a similar sense as in \citet{meng1994multiple}, then $\beta$ should be a function of $\theta$; that is, $\beta=h(\theta)$ and $\beta^{*}=h(\theta^{*})$. By the continuous mapping theorem, since $\hat{\theta}^{*}$ is consistent for $\theta$ as $n\rightarrow\infty$ and $\epsilon\rightarrow\infty$, so is $\hat{\beta}^*$ for $\beta$. With $m$ set of synthetic data, if $\hat{\beta}^{*(i)}$ is consistent for $\beta$ for $i=1,\ldots,m$, so is $m^{-1}\sum_{i=1}^m \beta^{*(i)}$ per the Slutsky’s theorem.

\textbf{Part b).} The proof is based in a similar framework as in \citet{MIbook} (inferences from multiple imputation) and \citet{reiter2003} (inferences from partial sample synthesis without sanitization), with necessary modifications to allow for the extra variability introduced during the sanitization of the posterior distribution before sampling. We first provide s a Bayesian derivation of the inference and then list the conditions under which
these inferences are valid from a frequentist perspective.

In the Bayesian framework,  the posterior variance of $\beta$ given synthetic data $\tilde{\x}^{*(i)}$ for $i\!=\!1,\ldots,m$ is
\begin{align}\vspace{-3pt}
&\V(\beta|\tilde{\x}^{*(1)},\ldots, \tilde{\x}^{*(m)})
=\V(\E(\beta|\x)|\tilde{\x}^{*(1)},\ldots, \tilde{\x}^{*(m)})\!+\!\E(\V(\beta|\x)|\tilde{\x}^{*(1)},\ldots, \tilde{\x}^{*(m)})\nonumber\\
=& \V(\hat{\beta}|\tilde{\x}^{*(1)},\ldots, \tilde{\x}^{*(m)})+ \E(\hat{v}|\tilde{\x}^{*(1)},\ldots, \tilde{\x}^{*(m)}),\label{appeqn:varcomb}
%=&\V(\hat{\beta}|\hat{\beta}^{*(1)},\ldots,\hat{\beta}^{*(m)})+ \E(v(\x)|v(\tilde{\x}^{*(1)}),\ldots, v(\tilde{\x}^{*(m)})),\label{appeqn:varcomb}
\vspace{-6pt}\end{align}
where $\hat{\beta}$  and $\hat{v}$ are the posterior mean and variance  of $\beta$, respectively, given the original data $\x$; $\hat{\beta}^{*(i)}$ and $\hat{v}^{*(i)}$ are the posterior mean and variance of $\beta^{*(i)}$, respectively, given $\tilde{\x}^{*(i)}$. By the large-sample theory, as $n\rightarrow\infty$,
\begin{eqnarray}
\beta |\x & \sim& N(\hat{\beta}, \hat{v})\label{eqn:app-post.theta}\\
\beta^{*(i)}|\x^{*(i)}& \sim& N(\hat{\beta}^{*(i)},\hat{v}^{*(i)}). \label{eqn:app-post.theta*1}
%\beta^{*(i)}|\tilde{\x}^{*(i)}& \sim& N(g(\tilde{\x}^{*(i)}), \hat{v}^{*(i)})).\label{eqn:app-post.theta*20}
\end{eqnarray}
Since $\beta^{*(i)}$  is independent for $i=1,\ldots,m$ conditional on $\tilde{x}^{*(i)}$, we have, from Eq.~(\ref{eqn:app-post.theta*1}), 
\begin{eqnarray}
\textstyle m^{-1}\sum_{i=1}^m\beta^{*(i)}|\tilde{\x}^{*(i)}&\sim& \textstyle N(m^{-1}\sum_{i=1}^m \hat{\beta}^{*(i)},   m^{-2}\sum_{i=1}^m\hat{v}^{*(i)}).\label{eqn:app-post.theta*2}
\end{eqnarray}
Since $f(\beta^*|\x)\xrightarrow[]{d}f(\beta|\x)$, per the Lyapunov CLT, we have, as  $m\!\rightarrow\!\infty$ 
\begin{equation}
\beta|\beta^{*(1)},\ldots,\beta^{*(m)} \sim N(\textstyle m^{-1}\sum_{i=1}^m\beta^{*(i)}, m^{-2}\sum_{j=1}^mv(\beta^{*(i)})) \label{eqn:app-L1}
 \end{equation}
Eqs (\ref{eqn:app-post.theta}), (\ref{eqn:app-post.theta*2}), and (\ref{eqn:app-L1}) taken together, it suggests
\begin{equation}
\hat{\beta}|\tilde{\x}^{*(1)},\ldots,\tilde{\x}^{*(m)}\sim N(\textstyle m^{-1}\sum_{i=1}^m\hat{\beta}^{*(i)},m^{-2}\sum_{i=1}^m(v(\beta^{*(i)})+\hat{v}^{*(i)})).\label{eqn:between}
\end{equation}
In a similar manner, we obtain the conditional distribution of $v(\x)$ given $\tilde{\x}^{*(1)}, \ldots, \tilde{\x}^{*(m)}$
\begin{equation}
v(\x)|\tilde{\x}^{*(1)}, \ldots, \tilde{\x}^{*(m)}\!\sim\!\textstyle N( m^{-1}\sum_{i=1}^m\hat{v}^{*(i)}, m^{-2}\sum_{i=1}^m(v_v(\beta^{*(i)})+\hat{v}^{*(i)})). \label{eqn:within}
\end{equation}
Replace the two terms in Eq (\ref{appeqn:varcomb}) with the conditional variance from Eq (\ref{eqn:between}) and mean from Eq (\ref{eqn:within}) and denote $m^{-1}\sum_{i=1}^m\left(v(\beta^{*(i)})+\hat{v}^{*(i)})\right)$ by $b$ and $m^{-1}\sum_{i=1}^m\hat{v}^{*(i)})$ by $\varpi$, then
$$\V(\beta|\tilde{\x}^{*(1)},\ldots, \tilde{\x}^{*(m)}) =  \textstyle m^{-1}b+\varpi.$$
For finite $m$, $b$ is approximated by $(m-1)^{-1}\sum_{i=1}^m (\hat{\beta}^{*(i)}-\bar{\beta}^*)^2$ and $\varpi$ by $m^{-1}\sum_{i=1}^m\hat{v}^{*(i)}$.

Similar to \citet{reiter2003}, the regularity condition for $m^{-1}b+\varpi$ to be an asymptotically unbiased estimator for $\bar{\beta}^*$ in the frequentist framework are 1) $\E\big(\hat{\beta}^{*(i)}|\x\big)\rightarrow \hat{\beta}$; 2) $\E\big(m^{-1}\sum_{i=1}^m\hat{v}^{*(i)}|\x\big)\!\rightarrow\!\hat{v}$; $\E\big((m\!-\!1)^{-1}\sum_{i=1}^m  (\hat{\beta}^{*(i)}\!-\!\bar{\beta}^*)^2|\x\big)\!\rightarrow\! \mbox{V}(\beta^{*(i)}|\x)$.
% as $\epsilon\rightarrow\infty$ and $n\rightarrow\infty$;

\textbf{Part c).} The results in parts a) and b) suggest $\beta|\tilde{\x}^{*(1)},\ldots, \tilde{\x}^{*(m)}\sim N(\bar{\beta}^*, m^{-1}b+\varpi)$. For a finite $m$, the distribution can be obtained in a similar manner as in  \citet{reiter2003}, which is  $f(\beta|\tilde{\x}^{*(1)},\ldots, \tilde{\x}^{*(m)})\sim t_{\nu}(\bar{\beta}^{*}, m^{-1}b+\varpi)$ with $\nu =(m-1)(1+m\varpi/b)^2$.

\section{\large{Proof of Remark \ref{rem:w}}}\label{app:conjind}\vspace{-6pt}
\textbf{Part a).} The scale parameter of the Laplace distribution in the communal sanitization is $\lambda=\delta_{\s}\epsilon^{-1}= r\bar{\delta}_{\s}\epsilon^{-1}$, where $\bar{\delta}_{\s}$ is the average GS. When $w_i\equiv r^{-1}$, every statistic receives the same amount of budget $\epsilon/r$ in the individualized sanitization , and the scale parameter of the Laplace distribution for $s_i$  is $\lambda'=\delta_i(\epsilon w_j)^{-1}=r\delta_i\epsilon^{-1}$, which is $<\lambda$ if $\delta_i<\bar{\delta}_{\s}$, and $>\lambda$ if $\delta_i>\bar{\delta}_{\s}$.
\textbf{Part b).} The scale parameter for  the Laplace distribution for $s_i$ in the individualized sanitization is $\delta_i(\epsilon w_i)^{-1} =\epsilon^{-1}\delta_{s_i}(\delta_{s_i})^{-1}\!\sum_{i=1}^r\! \delta_i=\epsilon^{-1}\sum_{j=1}^r \delta_i$, which is the same as the scale parameter for  the Laplace distribution in the communal sanitization.

\end{document}